\documentclass[iop]{emulateapj}
\usepackage{apjfonts}

\usepackage{graphicx}
\usepackage{wasysym}

\shorttitle{BH spin origin in Galactic LMXBs}
\shortauthors{T. Fragos et al.}

\begin{document}

\title{The Origin of Black Hole Spin in Galactic Low-Mass X-ray Binaries }

\author{T.\ Fragos$^{1}$, J.\ E.\ McClintock$^{2}$} 
\altaffiltext{1}{Geneva Observatory, University of Geneva, Chemin des Maillettes 51, 1290 Sauverny, Switzerland}
\altaffiltext{2}{Harvard-Smithsonian Center for Astrophysics, 60 Garden Street, Cambridge, MA 02138 USA}
\email{anastasios.fragkos@unige.ch (corresponding author)}

\begin{abstract}

  Galactic field black hole (BH) low-mass X-ray binaries (LMXBs) are believed to form in situ
  via the evolution of isolated binaries. In the standard formation channel, these systems
  survived a common envelope phase, after which the remaining helium core of the primary star
  and the subsequently formed BH are not expected to be highly spinning. However, the measured
  spins of BHs in LMXBs cover the whole range of spin parameters. We propose here that the BH
  spin in LMXBs is acquired through accretion onto the BH after its formation. In order to test
  this hypothesis, we calculated extensive grids of detailed binary mass-transfer sequences. For
  each sequence, we examined whether, at any point in time, the calculated binary properties are
  in agreement with their observationally inferred counterparts of 16 Galactic LMXBs. The
  ``successful'' sequences give estimates of the mass that the BH has accreted since the onset
  of Roche-Lobe overflow. We find that in \emph{all} Galactic LMXBs with measured BH spin, the
  origin of the spin can be accounted for by the accreted matter, and we make predictions about
  the maximum BH spin in LMXBs where no measurement is yet available. Furthermore, we derive
  limits on the maximum spin that any BH can have depending on current properties of the binary
  it resides in. Finally we discuss the implication that our findings have on the BH birth-mass
  distribution, which is shifted by $\sim 1.5\,\rm M_{\odot}$ towards lower masses, compared to
  the currently observed one.

\end{abstract}

\keywords{black hole physics, Galaxy: stellar content, stars: binaries: close, stars: black holes, stars: evolution, X-rays: binaries}

\maketitle

\section{INTRODUCTION}

Stellar-mass black holes (BH) are the evolutionary remnants of massive stars ($\gtrsim 20\,\rm M_{\odot}$) \citep[e.g.][]{Georgy2009,Belczynski2010}. The existence of BHs is one of the most robust predictions in Einstein's theory of General Relativity. BHs can be fully described by three numbers: their mass, their spin (angular momentum), and their electric charge. Astrophysical BHs are believed to have negligible electric charge, so one is left with only two properties. Yet simply finding an isolated BH, much less measuring its properties of mass or spin, can be difficult.

Interacting binaries are arguably one of the most important astrophysical laboratories available for the study of compact objects, especially BHs. Accretion of matter from a close binary companion gives rise to X-ray emission and rejuvenates compact objects, rendering them detectable throughout the Galaxy and beyond. While some clues on the astrophysics of these X-Ray Binaries (XRBs) can be obtained from observations and modeling of their present-day properties, more comprehensive insight requires understanding their origin and evolutionary links to other stellar systems. 

Observations in 1972 of the XRB Cygnus X-1 provided the first strong evidence that BHs exist \citep{WM1972,Bolton1972}. Today, a total of 23 such XRB systems are known to contain a compact object too massive to be a neutron star or a degenerate star of any kind \citep[i.e. $M>3\,\rm M_{\odot}$;][]{Ozel2010, Farr2011}. The host systems of all known stellar-mass BHs are XRBs, i.e. mass-exchange binaries containing a non-degenerate star that supplies gas to the BH via a stellar wind or via Roche-lobe overflow (RLO) in a stream that emanates from the inner Lagrangian point. 

\subsection{Measuring the Spin of Accreting BHs}

\begin{deluxetable*}{lllll}
\tabletypesize{\scriptsize}
\tablewidth{0pt}
\tablecaption{\label{spin_table} Spin measurements results to date for nine stellar-mass BHs using the continuum-fitting method \tablenotemark{a}}
\tablehead{\colhead{Source} & \colhead{MT Type\tablenotemark{b}} & \colhead{$P_{\rm orb}$ (days)\tablenotemark{b}} & \colhead{Spin $a_*$}& \colhead{Reference}}
\startdata
GRS 1915+105&    RLO	&	33.9	&$>0.98$&            \citet{McClintock2006} \\
Cyg X-1&	 Wind	&	5.60	&$>0.983$&             \citet{Gou2014} \\
LMC X--1&        Wind	&	3.91	&$0.92_{-0.07}^{+0.05}$&   \citet{Gou2009} \\
M33 X--7&        Wind	&	3.45	&$0.84\pm0.05$&            \citet{Liu2008,Liu2010} \\ 
4U 1543--47&     RLO	&	1.12	&$0.80\pm0.05$&            \citet{Shafee2006} \\
GRO J1655--40&   RLO	&	2.62	&$0.70\pm0.05$&            \citet{Shafee2006} \\
XTE J1550--564&  RLO	&	1.54	&$0.34_{-0.28}^{+0.20}$&   \citet{Steiner2011} \\
LMC X--3&        RLO	&	1.70	&$0.25_{-0.16}^{+0.13}$&   \citet{Steiner:2014dx} \\
A0620--00&       RLO	&	0.32	&$0.12\pm0.19$&            \citet{Gou2010} \\
\enddata
\tablenotetext{a}{Errors are quoted at the 68\% level of confidence.}
\tablenotetext{b}{\citet{McCR2006} and references therein}
\end{deluxetable*}


\begin{figure}[b]
\centering
\includegraphics[width=0.49\textwidth]{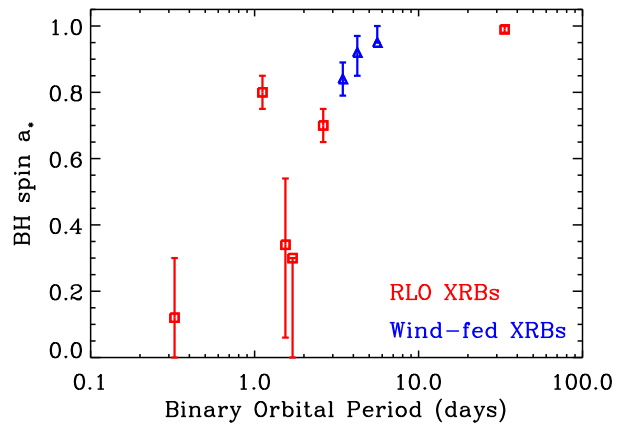}
\caption{\label{spin_plot} The spin parameter $a_*$ as a function of the binary orbital period, where persistent, wind-fed XRBs systems are plotted as (blue) triangles and transient, RLO XRBs as (red) squares, for the nine BH XRBs of Table~1 with measured BH spin using the continuum-fitting method.}
\end{figure}

Although the existence of stellar-mass BHs was confirmed several decades
ago via the dynamical measurement of their mass, the first attempt to
measure their spins was made much more recently \citep{Zhang1997}, and the 
first plausibly reliable results were obtained less than
a decade ago \citep{Shafee2006}.

To date, there are three methods that have been widely applied in estimating the spins of stellar-mass BHs \citep{RMcC2006}, namely, fitting the thermal continuum spectrum of the accretion disk, modeling the disk reflection spectrum with a focus on the Fe K line, and modeling high-frequency ($\sim100-450$~Hz) quasi-periodic oscillations (HFQPOs). While there are well-established models underpinning the first two methods, there is no agreed upon, or even leading, model of HFQPOs. Many classes of models have been proposed including several types of resonance models; global oscillation (``diskoseismic'') modes of the accretion disk; orbiting hot spots; tidal disruption of large inhomogeneities in the accretion flow; and the ``relativistic precession'' model, for which interesting results for two BHs were reported recently \citep{Motta2014, Motta2014b}. For discussion, critiques and references concerning these and other models, see, e.g., \citet{vdKlis2006, McCR2006, RM2009,Torok2011,DB2014}.

Presently, none of the models of HFQPOs is strongly preferred. Meanwhile, all of the models are basically dynamical and lack radiation mechanisms, and they largely fail to consider the established spectral properties of HFQPOs \citep{RMcC2006}. At present, an additional obstacle to attempting to use HFQPOs to validate a particular model and to estimate BH spin is the faintness of these transient oscillations and the paucity of data \citep{RMcC2006, Belloni2012}.

We turn now to considering the other two methods, which are generally referred to as the continuum-fitting method and the Fe-line method. The great importance of the Fe-line method is its dominant role in measuring the spins of supermassive BHs in AGN. In the Fe-line method, one determines the radius of the innermost stable circular orbit $R_{\rm ISCO}$, and hence the BH spin parameter\footnote {$a_* \equiv cJ/GM^2$ with $|a_*| \le 1$, where $M$ and $J$ are respectively the BH mass and angular momentum.}~by modeling the profile of the broad and skewed line that is formed in the inner disk by Doppler effects, light bending and gravitational redshift \citep{Fabian1989,Reynolds2013}. The line is the most prominent and easily-observed feature in the ``reflection'' spectrum, which is generated in a disk that is irradiated by a Compton power-law component.

In applying the continuum-fitting method, one fits the thermal continuum spectrum of a BH's accretion disk to the relativistic thin-disk model of \citet{NT1973} and thereby determines the radius of the inner edge of the disk \citep{McClintock2013}. One then identifies this radius with the radius of the innermost stable circular orbit $R_{\rm ISCO}$, which is simply related to the spin parameter $a_*$ for a BH of known mass \citep{Bardeen1972}. The method is simple: It is strictly analogous to measuring the radius of a star whose flux, temperature and distance are known. For this method to succeed, it is essential to have accurate estimates of BH mass $M_{\rm BH}$, disk inclination $i$ and source distance $D$.

In this paper we only use spin data derived via the continuum-fitting method, which for stellar-mass BHs we argue is the gold standard because of the relative virtues of this method: The thin-disk model is the simplest and most well-established model in strong-gravity accretion physics \citep{SS1973,NT1973}, and the model has been validated via general relativistic magnetohydrodynamic simulations \citep{Shafee2008,Penna2010,Kulkarni2011,Noble2011,Zhu2012}. There is a great abundance of suitable spectral data because a wide range of detectors are capable of providing such data ({\it RXTE} PCA, {\it Ginga} LAC, {\it ASCA} GIS, etc.), and most BHs remain for months in a disk-dominated state of moderate luminosity that is well-described by the thin-disk model. Finally, the problem of systematic errors (apart from the question of spin/orbit alignment) has been thoroughly addressed \citep{McClintock2013}.

By comparison, the available Fe-line spin data for stellar BHs is sparse and usually suffers from pileup effects, and the signal is faint relative to the continuum. The model is necessarily more complex than that of a thin, thermal disk because a reflection spectrum that is suitable for measuring spin requires that the source be in a strongly Comptonized (hard, steep power-law, or intermediate) state \citep{RMcC2006}. Several sources of systematic error have not yet been adequately explored, such as those associated with the assumptions that: in the hard state the disk's inner edge is at the ISCO; the disk has a constant-density atmosphere and can be described by a single state of ionization; the reflection models capture all the essential atomic physics.

Recently, \citet{NMcC2012} reported observational evidence for a correlation between jet power and BH spin. More specifically, they showed that the 5~GHz radio flux of transient ballistic jets in BH XRBs scales as the square of the BH spin parameter $a_*$ estimated via the continuum-fitting method. This is the first direct evidence that jets may be powered by BH spin energy. The evidence is still controversial, largely because of the small sample of sources \citep{RGF2013,McClintock2013}. \citet{Steiner2013} used the correlation between jet power and spin, and published radio and X-ray light-curve data, to estimate the spins of six other BH LMXBs.

The spins of ten stellar-mass BHs in XRBs of the Milky Way or nearby
galaxies have been measured using the continuum-fitting method. Three of
these systems are persistent, wind-fed high-mass XRBs (HMXBs) and the
remaining seven are transient, Roche lobe overfilling low-mass XRBs
(LMXBs)\footnote{In this paper, we disregard one of these LMXBs, namely
  H1743--322 \citep{Steiner2012}, because its orbital period is not
  known}. Table 1 lists the dimensionless spin parameter $a_*$, the
mass-transfer (MT) type, and the orbital period ($P_{\rm orb}$) of the
nine BH-XRBs. Figure 1 shows the spin parameter $a_*$ as a function of
the orbital period of the binary, where persistent systems are plotted
with triangles and transient RLO systems with squares.  It is evident
that all three HMXBs with massive O-star companions contain a highly
spinning BH ($a_∗ > 0.8$), while the spins of transient BHs span the
entire range of prograde values from near-zero (e.g. A0620--00) to
near-maximal (e.g. GRS 1915+105).

We conclude by noting that Morningstar et al.\ (2014) recently reported a
retrograde spin for Nova Mus 1991 (GRS 1124--683). We exclude this
result from Table~1 and do not use it because it is based on an
approximate analysis of the X-ray data (e.g. the treatment of the
effects of spectral hardening is crude), and on provisional literature
estimates of $M$, $i$ and $D$ \citep[e.g., $M$ is reported in a
non-refereed conference paper by][]{Gelino2004}.

\subsection{Early Theoretical Efforts}

Undoubtedly, the internal differential rotation of massive stars, which are the progenitors of stellar-mass BHs, plays a crucial role in the understanding of the origin of BH spin. Stellar rotation has been a subject of intense study for over a decade now \citep[][and references therein]{MM2012,Langer2012}. Although significant advances have been achieved in the codes used to study the effects of rotation on stellar structure and evolution, the basic physics of the angular momentum transport mechanisms are still uncertain \citep[e.g.][]{Kawaler1988, HWS2005, Suijs2008}. Despite these uncertainties, calibrating stellar models using observed rotation rates of young pulsars and white dwarves, and taking into account the most recent stellar wind estimates for massive stars and their dependence on metallicity, it is becoming widely accepted that evolutionary models of single stars fail to predict the existence of highly spinning BHs at solar-like metallicity, like those observed in some Galactic XRBs \citep[e.g.][]{WB2006,Meynet2008,YLN2006}. A recent astroseismic study \citep{Beck2012} reported that the cores of three red-giant stars rotate about 10 times faster than their surfaces, indicating that the angular momentum transport in the interior of stars must be relatively efficient. For comparison, stellar models with no additional angular momentum transport mechanisms predict that the core of a red giant star completely decouples from the envelope and can be rotating as much as 1000 times faster than its surface \citep{MM2012}.

Before the first stellar-mass BH spin measurements were made,
\citet{LBW2002} attempted to predict the spin parameter $a_*$ based on
the current binary properties of the host system. They targeted mainly a
subclass of LMXBs for which magnetic braking does not operate, namely,
systems where the RLO donor is a sub-giant or giant star with mass
$\gtrsim 1.5\,\rm M_{\odot}$. Using simplistic, order-of-magnitude
arguments about the evolutionary history of each of the BH LMXBs, they
were able to estimate the pre-supernova period of the binary. Finally,
they made the crucial assumptions that: (i) the pre-supernova binary was
fully synchronized during the preceding common-envelope phase and that
the helium star rotated as a solid body until the core had collapsed,
and (ii) that there was no mass loss or asymmetries during the
core-collapse itself. This analysis allowed them to estimate the birth
spin parameter $a_*$ of the BH, which they also assumed to be equal to
the present-day value. Although the very first two measurements of BH
spin, those for GRO\,J1655-40 and 4U\,1543-47 \citep{Shafee2006}, were
consistent with the predictions of \citet{LBW2002}, subsequent spin
measurements for other LMXBs with both tighter and wider orbits
disproved their model.  In fact, based on the currently available BH
spin measurements, one infers a positive correlation between the BH spin
and the orbital period of the binaries they reside in (see
Figure~\ref{spin_plot}), which is the opposite of what the analysis of
\citet{LBW2002} suggest \citep[see Figures 10-12 in ][]{LBW2002}.


In a series of papers, \citet{Mendez2008,Mendez2011a} and
\citet{Mendez2011b} claimed that case-C MT (MT while the donor star is
in the helium burning phase) or case-M evolution (which involves
tidally-locked, rotationally-mixed, chemically-homogeneous stars in a
close binary), cannot explain the observed BH spins in HMXBs such as LMC
X-1 or M33 X-7. The authors claim that for any such system and any
evolutionary scenario that the spin of the BH results from hypercritical
mass accretion that occurs after the formation of the BH. This analysis
has three major drawbacks: \emph{(i)} It is unclear what mass reservoir
could feed the BH at a high enough accretion rate in order to achieve
hypercritical accretion. The stellar wind of the O-star companion in
these short-lived systems cannot provide the necessary mass to spin up
the BH to high $a_*$ (see Figure~\ref{Macc_aspin}). \emph{(ii)} Even if
the necessary MT rate was somehow achieved, the MT would be dynamically
unstable and would lead to the merger of the binary
\citep{Valsecchi2010}. \emph{(iii)} If an episode of hypercritical MT
was initiated after the BH formed, then this accretion has to be
continued until the present day. However, there is no observational
indication in any of the observed systems that this type of accretion is
currently ongoing.

\section{Angular Momentum Gain in Stellar-Mass BHs due to Long-Term Stable Accretion}

Galactic field LMXBs, like those for which BH spin measurements are
available, are believed to form in situ via the evolution of isolated
binaries. The standard formation channel \citep{BvdH1991, TvdH2006}
involves a primordial binary system with a large mass ratio; the more
massive star evolves quickly to the giant branch and the system goes
into a common envelope phase. During this phase, the orbit of the system
changes dramatically, as orbital energy is lost due to friction between
the unevolved star and the envelope of the giant. Part of this
orbital energy is used to expel the envelope of the giant star. The
common envelope phase results in a tighter binary system with an
unevolved low-mass main-sequence star orbiting around the core of the
massive star. Soon, the massive core collapses to form a compact
object. If the binary does not get disrupted or merge in any of the
stages described above, angular momentum loss mechanisms, such as
magnetic braking, can further shrink the orbit. The companion star
eventually overflows its Roche lobe, transferring mass onto the compact
object and initiating the system's X-ray phase, which lasts $\sim 1\,\rm
Gyr$; up to a few solar masses of material can be accreted onto the BH
during this phase.

At the onset of the common envelope phase, the primary star, which is
the BH progenitor, has expanded to a typical radius of $100-1000\,\rm
R_{\odot}$. Up to that moment, the expansion of the star, the stellar
wind mass loss, and the tidal interactions with the companion star that
tend to synchronize the rotation of the primary with the wide orbit
will, most probably, carry away any significant initial angular momentum
that the primary had, and thus spin it down to low rotation
rates. During the common envelope phase itself, while the orbit is
shrinking significantly, the short timescale (the common envelope is
expected to last for only $\lesssim1$ thermal timescale) and the break of
corotation of the binary will not allow any significant transfer of
angular momentum from the orbit to the core of the primary star
\citep[e.g.][]{IPS2002,TR2010,Ivanova2012b}. Hence, the helium core is
expected to be spinning relatively slowly, with a rotational velocity
similar to that the primary star had just prior to the onset of the
common envelope.

The common-envelope evolution is followed by a phase of detached
evolution with the still unevolved low-mass secondary star and the
helium core of the primary star in a close orbit. The tides are expected
to operate efficiently in this phase and bring the binary in
synchronization in thousands or tens of thousands of years
\citep{vdHY2007}, which is much sorter than the typical lifetimes of
helium stars of $\sim 10\,\rm M_{sun}$ \citep{Paczynski1971}. Hence, the
spins of these helium stars will be most likely fully synchronized with
their orbits during their core-helium burning phase
\citep{Izzard2004,Podsiadlowski2004}. Whether tidal synchronization will
lead to further shrinkage of the orbit and spin up of the BH progenitor
star is uncertain, and it depends strongly on the metallicity of the
system and its initial post-common-envelope orbital
period. \citet{Detmers2008} studied in detail the tidal spin up of a
helium-star BH progenitor, that is in a tight orbit with a neutron star.  They found
that even in this extreme configuration, where the companion is a
neutron star and can fit in an orbit of a few days, the helium star
cannot be spun up enough to produce a collapsar and hence a highly
spinning BH. Although the synchronization of a helium star's
spin to the orbit happens quickly, the whole process is limited by the
widening of the binary orbit induced by the strong Wolf-Rayet wind or by
the radius-evolution of the Wolf-Rayet star, which most often leads to a
binary merger. The situation only gets worse if one considers
non-degenerate, hydrogen-rich, low-to-intermediate mass stellar
companions, which have much larger radii, because mergers in this case
are at longer orbital periods.


As a consequence, the BHs formed in these systems are more likely to have low birth spin. However, the measured spins of BHs in LMXBs cover the whole range of spin parameters from $a_*=0$ to almost $a_*=1$. If the assumptions above are even approximately valid, then this implies that the BH spin in LMXBs is determined by the matter that the BH has accreted during the long stable accretion phase of the system. Support for this view is provided by \citet{PRH2003} who have shown that for certain initial binary configurations, and depending on the assumptions for the accretion efficiency, an initial non-spinning BH can be spun up to $a_*>0.8$ by material accreted from a RLO companion star.
 
\begin{figure}
\centering
\includegraphics[width=0.5\textwidth]{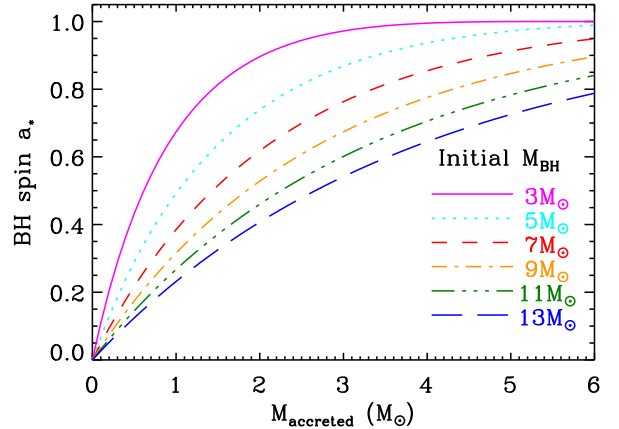}
\caption{\label{Macc_aspin} The dimensionless spin parameter $a_*$ as a function of the amount of matter accreted onto a BH, for different birth BH masses. The birth spin of the BH is assumed to be zero and the matter accreted is assumed to carry the specific angular momentum of the innermost stable circular orbit \citep{Thorne1974}.}
\end{figure}

In order to test this hypothesis, we study the evolutionary history of
each Galactic BH LMXBs with measured or estimated BH spin, following a
methodology similar to \citet{Willems2005} and
\citet{Fragos2009b}. Since direct backwards integration of the
differential equations governing stellar and binary evolution is not
feasible (nor unique), reversing the stellar and binary evolution
requires the calculation of extensive grids of evolutionary sequences
for binaries in which a BH accretes matter from a close companion. For
each evolutionary sequence, we examine whether, at any point in time,
the calculated binary properties (orbital period $P_{\rm orb}$; BH mass
$M_{\rm BH}$; donor mass $M_{2}$; donor's effective temperature $T_{\rm
  eff}$; and MT rate) are in agreement with their observationally
inferred counterparts. While many sequences are able to satisfy some of
the observational constraints, only a finite set satisfy all of them
simultaneously. MT sequences that simultaneously satisfy all
observational constraints represent possible progenitors of the
considered LMXB and thus yield possible donor and BH masses, and orbital
periods at the onset of the MT phase. But most importantly for our work,
these sequences give estimates of the total amount of matter that the BH
has accreted from the onset of RLO until today. Figure~\ref{Macc_aspin}
shows the dimensionless spin parameter $a_*$ as a function of the total
mass accreted onto the BH, for different birth BH masses. The birth spin
of the BH is assumed to be zero and the matter accreted is assumed to
carry the specific angular momentum of the innermost stable circular
orbit \citep{Thorne1974}. Based on the constraints we derive on the
total amount of matter that the BH has accreted, for each of the
considered LMXBs we are able to estimate the expected spin of the BH.


\section{Observational Sample}

\begin{deluxetable*}{ccccccccc}
\tablecolumns{9}
\tabletypesize{\scriptsize}
\tablecaption{Currently observed properties of 16 Galactic LMXBs with a dynamically confirmed BH and at least an orbital period measurement.    
\label{ObsProp}}
\tablehead{ 
   \colhead{Symbol} & 
   \colhead{System Name} &
   \colhead{$M_{\rm BH}\,[M_\odot]$\tablenotemark{a}} & 
   \colhead{$M_2\,[M_\odot]$\tablenotemark{b}} & 
   \colhead{$P_{\rm orb}\,[days]$\tablenotemark{c}} &
   \colhead{Spectral type\tablenotemark{d}} &
   \colhead{$T_{\rm Eff}\,[K]$\tablenotemark{e}} &
   \colhead{$a_*$\tablenotemark{f}} &
   \colhead{References\tablenotemark{g}} 
     }
\startdata
$\blacksquare$			&	GRS 1915+105	    &	$12.4\pm2.0$			&	$0.58\pm0.33$	&	33.85	&	K0III-K3III	&	$4100 - 5433^{\star}$   &	$0.95\pm0.05$			&	1, 2, 3, 4, 5, 6 \\ 
$\blacktriangledown$	&	4U 1543-47		    &	$9.4\pm2.0$				&	$2.7\pm1.0$		&	1.116	&	A2V			&	$9000\pm 500$		    &	$0.8\pm0.1$				&	7, 8, 9, 10\\ 
$\blacktriangle$		&	GRO J1655-40$^{GBO}$\tablenotemark{*}&	$6.3\pm0.5$				&	$2.4\pm0.4$		&	2.622	&	F5III-F7III	&	$5706 - 6466^{\star}$	&	$0.7\pm0.1$				&	11, 12, 10\\ 
$\blacktriangleright$	&	GRO J1655-40$^{BP}$\tablenotemark{*} &	$5.4\pm0.3$				&	$1.45\pm0.35$	&	   '' 	&	  ''        &	      ''              	&	    ''        			&	13\\ 
$\blacklozenge$			&	XTE J1550-564	    &	$9.1\pm0.61$			&	$0.30\pm0.07$	&	1.542	&	K3III/V		&	$4700\pm 250$		    &	$0.34\pm0.2$			&	14, 15\\ 
$\CIRCLE$				&	A0620-00		    &	$6.61\pm0.25$			&	$0.40\pm0.05$	&	0.323	&	K5V-K7V		&	$3800 - 4910^{\star}$	&	$0.12\pm0.19$			&	16, 17, 18, 19\\ 
$\square$				&	GRS 1124-683	    &	$6.95\pm1.1$			&	$0.9\pm0.3$		&	0.433	&	K3V-K5V		&	$4065 - 5214^{\star}$	&	$0.25\pm0.15^{\dagger}$	&	20, 21, 22, 23\\ 
$\triangledown$			&	GX 339-4		    &	$8.0\pm1.0^{\ddagger}$	&	N/A				&	1.754	&	N/A			&	N/A						&	$0.25\pm0.15^{\dagger}$	&	24, 25, 23\\ 
$\triangle$				&	XTE J1859+226	    &	$8.0\pm1.0^{\ddagger}$	&	N/A				&	0.383	&	N/A			&	N/A						&	$0.25\pm0.15^{\dagger}$	&	26, 27, 23\\ 
$\Circle$				&	GS 2000+251		    &	$8.0\pm1.0^{\ddagger}$	&	$0.34\pm0.09$	&	0.344	&	K3V-K6V		&	$3915 - 5214^{\star}$	&	$0.05\pm0.05^{\dagger}$	&	23, 28, 29, 30, 31\\ 
						&	GRO J0422+32	    &	$8.0\pm1.0^{\ddagger}$	&	$0.95\pm0.25$	&	0.212	&	M1V-M4V		&	$2905 - 4378^{\star}$	&	--						&	32\\ 
						&	GRS 1009-45		    &	$8.5\pm1.0$				&	$0.54\pm0.10$	&	0.285	&	K7V-M0V		&	$3540 - 4640^{\star}$	&	--						&	33, 34\\ 
						&	GS 1354-64		    &	$8.0\pm1.0^{\ddagger}$	&	N/A				&	2.545	&	G0III-G5III	&	$4985 - 6097^{\star}$	&	--						&	35, 36\\ 
						&	GS 2023+338		    &	$9.0\pm0.6$				&	$0.54\pm0.05$	&	6.471	&	K0III-K3III	&	$4100 - 5433^{\star}$	&	--						&	37, 38, 39\\ 
						&	H1705-250		    &	$6.4\pm1.5$			    &	$0.25\pm0.17$   &	0.521	&	K3V-M0V		&	$3540 - 5214^{\star}$	&	--						&	40, 41\\ 
						&	V4641 Sgr		    &	$6.4\pm0.6$				&	$2.9\pm0.4$		&	2.817	&	B9III		&	$10500 \pm 200$			&	--						&	42, 43, 44\\ 
						&	XTE J1118+480	    &	$7.6\pm0.7$			    &	$0.18\pm0.07$	&	0.170	&	K7V-M1V		&	$3405 - 4640^{\star}$	&	--						&	45, 46, 47\\ 
\enddata
\tablenotetext{a}{BH mass}
\tablenotetext{b}{Donor star mass}
\tablenotetext{c}{Orbital period}
\tablenotetext{d}{Spectral type of the donor star}

\tablenotetext{e}{Donor star's effective temperature. When the value is denoted with a $\star$, $T_{\rm Eff}$ is not directly measured, but is instead inferred by the reported spectral type. All transformations of spectral types to effective temperatures were done based on the tables provided in \citet{Gray2008} and \citep{Cox2000}, assuming typical one-standard deviation uncertainties of: $\pm 1000\,\rm K$ for O9V-B2V stars, $\pm 250\,\rm K$ for B3V-B9V, $\pm 100\,\rm K$ for A0V-M6V, $\pm 100\,\rm K$ for F0III-F9III, $\pm 50\,\rm K$ for G0III-M5III, and $\pm 70\,\rm K$ for M0III-M6III \citep{Cox2000}. When there is a discrepancy between the values reported in \citet{Gray2008} and \citep{Cox2000}, the combined effective temperature range from the two tables is taken into account.}
\tablenotetext{f}{BH spin parameter. When the value is denoted with a $\dagger$, $a_*$ is not directly measured but estimated from the maximum jet power of the system  \citep{Steiner2013}}
\tablenotetext{$\ddagger$}{No reliable or accurate dynamical BH mass measurement is available for this system. Instead, the BH mass reported in this table is a fiducial value based on the derived BH mass distribution by \citet{Ozel2010}. For GX 339-4 this value is also consistent with a BH mass of $7.5\pm0.8\,\rm M_{\odot}$ reported by \citet{Chen2011} based on X-ray timing data. }
\tablenotetext{*}{\textbf{GBO,BP:} Two analyses for the determination of the BH and companion mass are available in the literature for GRO J1655-40, the one by \citet{GBO2001} denoted in the table by the superscript ``GBO'', and the one by \citet{BP2002} denoted by the superscript ``BP''. At present it is unclear which of the two analyses (GBO or BP) yields the most reliable estimates of $M_{\rm BH}$ and $M_2$. We therefore consider both sets of constraints and study them as two separate cases. }
\tablenotetext{g}{\textbf{References:} (1) \citet{Reid2014}, (2) \citet{Steeghs2013}, (3) \citet{Greiner2001}, (4) \citet{Greiner2001b}, (5) \citet{HG2004}, (6) \citet{McClintock2006}, (7) \citet{Orosz2003}, (8) Orosz et al.\ (in preparation), (9) \citet{Orosz1998}, (10) \citet{Shafee2006}, (11) \citet{GBO2001}, (12) \citet{Shahbaz1999}, (13) \citet{BP2002}, (14) \citet{Orosz2011}, (15) \citet{Steiner2011}, (16) \citet{Johannsen2009}, (17) \citet{Cantrell2010}, (18) \citet{NSV2008}, (19) \citet{Gou2010}, (20) \citet{Orosz1996}, (21) \citet{Gelino2001}, (22) \citet{Casares1997}, (23) \citet{Steiner2013}, (24) \citet{Cowley1987}, (25) \citet{Hynes2003}, (26) \citet{Zurita2002}, (27) \citet{FC2001}, (28) \citet{Harlaftis1996}, (29) \citet{Filippenko1995}, (30) \citet{Casares1995}, (31) \citet{Barret1996}, (32) \citet{Harlaftis1999}, (33) \citet{Filippenko1999}, (34) \citet{Macias2011},  (35) \citet{Casares2009}, (36) \citet{Casares2004}, (37) \citet{CC1994}, (38) \citet{Khargharia2010}, (39) \citet{Hynes2009}, (40) \citet{Remillard1996}, (41) \citet{Harlaftis1997}, (42) \citet{Orosz2001}, (43) \citet{Sadakane2006}, (44) \citet{MacDonald2014}, (45) \citet{Gonzalez2012}, (46) \citet{Khargharia2013}, (47) \citet{Calvelo2009}}

\end{deluxetable*}

For the rest of our analysis we will consider a sample of the 16
dynamically confirmed Galactic BH LMXBs for which the orbital period of
the binary is known. Table~\ref{ObsProp} summarizes our adopted values
of the observed properties of these systems, including the masses of the
two binary components, the effective temperature and the spectral type
of the donor star, the orbital period of the binary, and the BH spin
measurement, when available. Filled symbols correspond to systems for
which the BH spin has been measured using the continuum-fitting method,
and open symbols systems for which the BH spin has been estimated via
the jet power - BH spin correlation. Each symbol corresponds to the same
system in all figures for the rest of the paper. Finally, all reported
errors or plotted error bars correspond to one standard deviation,
unless otherwise specified.

In the case of GRO\,J1655-40, we present two analyses, each
corresponding to different and inconsistent values of the BH and
secondary mass that appear in the literature, since it is unclear which
results are more reliable.  In Table~\ref{ObsProp} and throughout the
rest of the paper, the pair of masses reported by \citet{GBO2001} for
GRO\,J1655-40 are denoted by the superscript ``GBO'', and those by
\citet{BP2002} are denoted by the superscript ``BP''.

We should note here that LMC X-3 is excluded from our observed comparison sample, despite being a transient RLO XRB with a BH accretor and having new robust observational constraints on its physical properties \citep{Orosz2014,Steiner2014}. The reason for this exclusion is that the metallicity of its donor star is significantly sub-solar ($Z\lesssim 0.4Z_\odot$), and thus a comparison of its observed properties with a grid of binary MT sequences run at solar metallicity (see Section 4) is not appropriate. A detailed analysis of the evolutionary history of LMC X-3 will follow in a separate paper.

\section{The grid of MT calculations}

We used the publicly available stellar evolution code {\tt MESA}
\citep[Modules for Experiments in Stellar Astrophysics][]{Paxton2011,
  Paxton2013} in order to calculate a grid of $\sim 28,000$ evolutionary
sequences for BH XRBs undergoing mass transfer. The sequences cover the
the available initial parameter space for the masses of the BH and the
donor star, and for the orbital period at the onset of the
RLO. Specifically we consider initial BH masses ($M_{\rm BH}$) from
$3\,\rm M_\odot$ to $10\,\rm M_\odot$ in increments of $1\,\rm M_\odot$;
donor star masses ($M_{2}$) between $0.5\,\rm M_\odot$ and $10.0\,\rm
M_\odot$ in increments of $0.1-0.2\,\rm M_\odot$; and initial binary
orbital periods between $0.2\,\rm days$ and $100.0\,\rm days$ in
increments of $0.05-5.0\,\rm days$. Since we assume that all systems
were formed from isolated primordial binaries in the Galactic disk, we
adopt a solar metallicity for the donor stars in our MT
calculations. For all the simulations we have used version 5527 of {\tt
  MESA} along with the {\tt MESA} Software Development Kit released on
09/18/2013, and we follow the implicit MT rate calculation
prescription\footnote{The detailed MESA input (inlist) files used in our
  simulations are available online at:
  $http://mesastar.org/results$}. Since the goal of this work is to test
the hypothesis that the measured values of a BH's spin can be accounted
for solely by the quantity of mass it accretes during its XRB phase, we
consider the limiting case of fully conservative MT, even when the
accretion rate exceeds the Eddington accretion rate ($\dot{M}_{\rm
  Edd}$). We do acknowledge that the accretion efficiency is a major
source of uncertainty in these systems, which is why the amount of
material accreted onto the BH in each of the MT sequences should only be
considered as an upper limit. Finally, we should mention that the MT
sequences were terminated when either of the following criteria were
satisfied: \emph{(i)} the age of the system exceeds the age of the
Universe ($13.7\,\rm Gyr$), \emph{(ii)} the orbital period exceeds one
year, \emph{(iii)} the mass of the donor star becomes less than $0.08
\,\rm M_{\odot}$, or \emph{(iv)} or the donor star becomes degenerate.

\begin{figure*}
\centering
\includegraphics[width=0.49\textwidth]{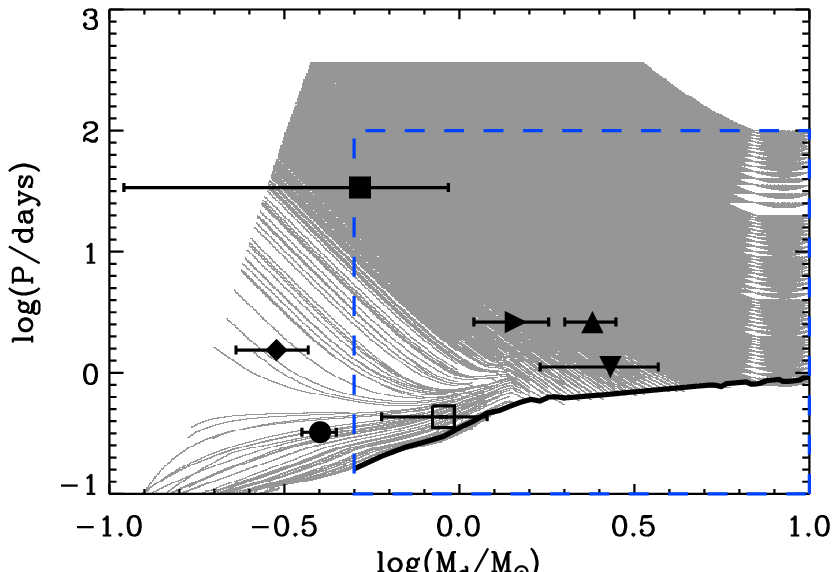}
\includegraphics[width=0.49\textwidth]{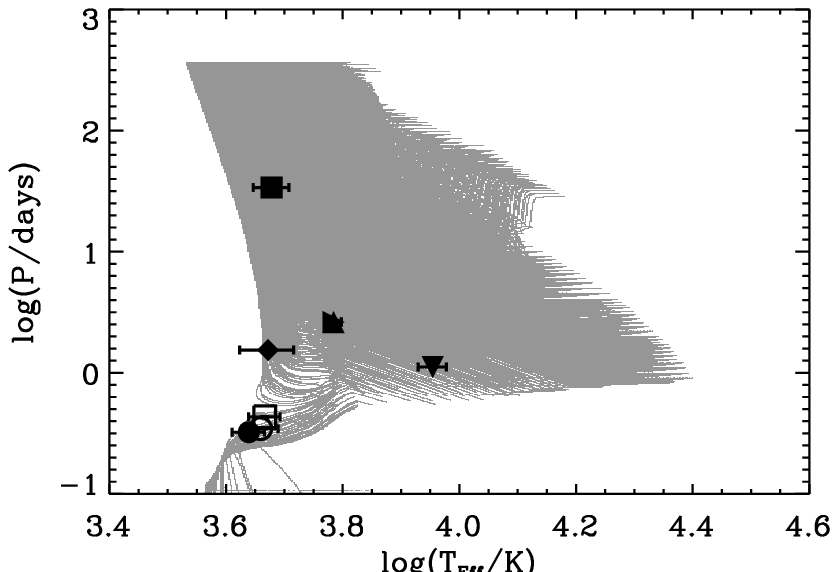}
\caption{\label{grid} The gray shaded area shows the parameter space in
  the orbital period-donor star's mass \textbf{(left panel)} and in the
  orbital period-donor star's effective temperature \textbf{(right
    panel)} plane that is covered by our grid of MT sequences, assuming
  an initial BH mass of $7\,\rm M_{\odot}$. The black solid line in the
  left panel shows the ZAMS mass-radius relation. No MT sequence track
  can cross to the area below this line. The area enclosed by the (blue)
  dashed line in the left panel shows the parameter space of initial
  conditions that is covered by our grid. For comparison we have
  over-plotted the orbital periods, donor star masses and effective
  temperatures of the 9 Galactic LMXBs with measured or estimated spins
  listed in Table~\ref{ObsProp}. }
\end{figure*}

Figure~\ref{grid} is a visual representation of the parameter space
covered by our grid of MT sequences, for an initial BH mass of $7\,\rm
M_\odot$. The region enclosed by the dashed line in the left panel
denotes the considered parameter space of initial conditions, donor mass
and orbital period at the onset of RLO. The gray shaded region shows the
parameter space, in the donor mass - orbital period plane and donor's
effective temperature - orbital period plane, that is covered by the
evolutionary tracks. Focusing on the donor mass - orbital period plane
one can see that the parameter space to be covered by our MT sequences
is finite. The black solid line shows the ZAMS mass-radius relation. No
MT sequence track can cross to the area below this line because the
initial mass would then be too great to fit within its Roche
lobe. Furthermore, we have adopted a maximum initial mass of $10\,\rm
M_{\odot}$ for the donor star and a maximum orbital period of one year;
if these limits are exceeded, the MT sequence is terminated. These two
limits create the sharp boundaries on the right and on the top side of
the gray shaded region respectively. Finally, the sharp boundary on the
left side of the gray region is a result of our criterion for
termination of a MT sequence when the donor star becomes
degenerate. This boundary coincides with the mass-radius relation of
helium white dwarves \citep{Rappaport1995}.

On the same figure we overlay the present-day observed properties of
seven out of the nine systems from Table~\ref{ObsProp} for which
measurements of the donor's mass and effective temperature, and a BH
spin measurement or estimate are available. We should note here that the
density of tracks shown in Figure~\ref{grid} is \emph{not} proportional
of the probability of observing a system in a specific part of the
parameter space. On the contrary, in the region near the bifurcation
period\footnote{The bifurcation period is the critical orbital period
  that separates the formation of converging systems (which evolve
  towards shorter orbital periods until the mass-losing component
  becomes degenerate and an ultra-compact binary is formed) from the
  formation of diverging systems (which evolve towards longer orbital
  periods until the mass-losing star has lost its envelope and a wide
  detached binary is formed). The exact position of the bifurcation
  period in the diagram depends on assumptions about the strength of
  the magnetic breaking and the accretion efficiency.} (lower left
corner of both panels), where the tracks diverge from each other, the
density of tracks is low. However, the systems evolve slowly along these
tracks which makes observing a system in this region very likely.  In
fact, half of the observed systems lie in this part of the
parameter space.

\section{Results}

\subsection{Unravelling the Evolutionary History of Galactic LMXBs back to the Onset of RLO}

For each of the calculated MT sequences of our grid, we checked if there is a time during the evolution of each system in Table~\ref{ObsProp} at which its properties satisfy simultaneously all the observational constraints: the masses of the BH and the donor star, the temperature of the donor star, and the orbital period. In all cases, we consider that a MT sequence satisfies an observational constraint when the calculated model property is within two standard deviations from the observed value. Since all LMXBs in Table~\ref{ObsProp} are X-ray transients, we also check whether the MT rate is below the critical rate for the occurrence of thermal disk instabilities, which are believed to cause the transient behavior of LMXBs \citep{vanParadijs1996,KKB1996,Dubus1999,MPH2002}.

MT sequences that simultaneously satisfy all observational constraints represent possible progenitors of the considered XRB and thus yield possible donor and BH masses and orbital periods at the onset of the MT phase. The time at which the sequences satisfy all present observational constraints provides us with an estimate for the age of the donor. In addition, this age directly gives us the age of the system and the time since the BH formed, assuming that the donor was approximately unevolved at that time. Finally, from each MT sequence that successfully reproduces all the observed properties of a given LMXB, we are able to derive the maximum amount of material that has been accreted onto the BH. This is a crucial quantity that allows us to calculate the predicted spin of the BH, assuming that the BH was initially not spinning and that the accreted material carries the specific angular momentum of the ISCO \citep[][ see also Figure~\ref{Macc_aspin}]{Thorne1974}. We again note that, since we assumed fully conservative MT, the estimated amount of accreted material, and hence the estimated BH spin, should only be thought of as upper limits.

\begin{figure*}
\centering
\includegraphics[width=0.49\textwidth]{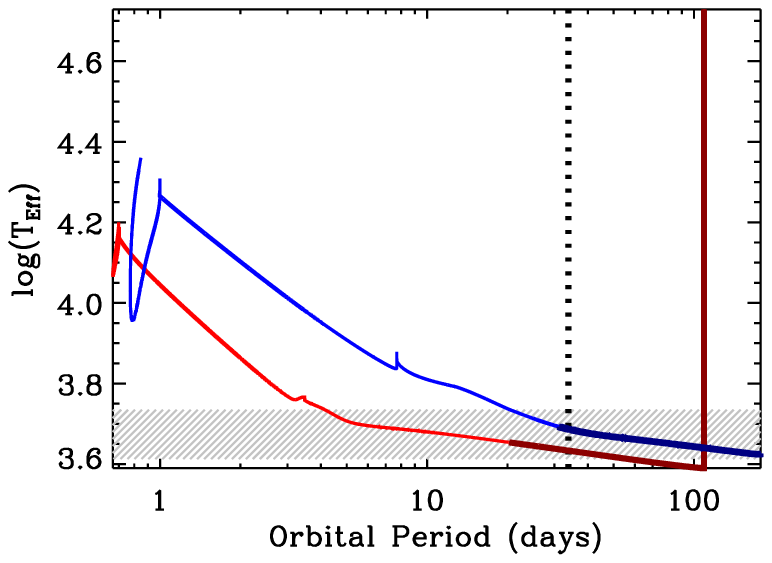}
\includegraphics[width=0.49\textwidth]{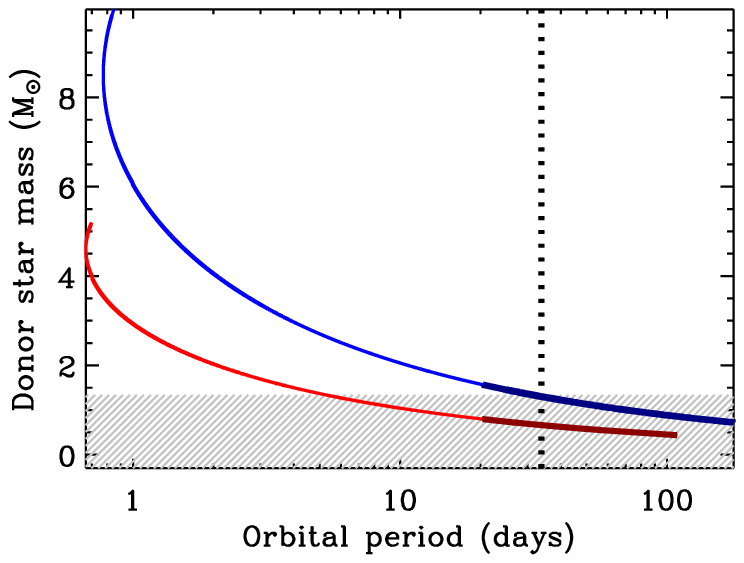}
\caption{\label{MTtracks} Evolutionary tracks of two selected MT sequences with different initial component masses and orbital periods that at some point during the evolution of the system satisfy simultaneously all the observational constraints of the Galactic LMXB GRS\,1915+105. The \emph{left panel} shows the variation of the orbital period as a function of the donor star's effective temperature, and the \emph{right panel} displays the variation of the donor mass as a function of the orbital period. The grey hatched regions indicate the observational constraint (at 90\% confidence) on the present-day donor's effective temperature (left panel) and mass (right panel). The vertical dotted line in both panels represents the currently observed orbital period of GRS\,1915+105. In both panels, the thick/darker part of the evolutionary tracks indicates the part of the sequence where the modeled system satisfies the observational constraints not shown in the corresponding panel (e.g. BH and donor mass, MT rate, and BH spin for the left panel). Successful sequences that simultaneously satisfy all observational constraints are therefore given by tracks for which the thick part of the track crosses the $P_{\rm orb}$ constraint line inside the hatched region on the left panel. In both panels the \emph{light grey (red) line} corresponds to a MT sequence with $M_2=5.2\,\rm M_{\odot}$, $M_{\rm BH}=4.0\,\rm M_{\odot}$, and $P_{\rm orb}=0.7\,\rm days$ at the start of RLO. The \emph{dark grey (blue) line} indicates a MT sequence with $M_2=10.0\rm\,M_{\odot}$, $M_{\rm BH}=7.0\,\rm M_{\odot}$, and $P_{\rm orb}=0.85\,\rm days$ at the start of RLO. The two selected MT sequences are the ones with the least massive and most massive donor star, respectively, that are able to satisfy at some point in their evolution all the observational constraints of GRS\,1915+105, including its very high spin.}
\end{figure*}

In Figure~\ref{MTtracks} we show, as an example, the systematic behavior
of two selected MT sequences with different initial component masses and
orbital periods. The two selected sequences satisfy at some point of
their evolution all the observational constraints for GRS\,1915+105,
including its very high BH spin. The light grey (red) line corresponds
to a MT sequence with $M_2=5.2\,\rm M_{\odot}$, $M_{\rm BH}=4.0\,\rm
M_{\odot}$, and $P_{\rm orb}=0.7\,\rm days$ at the start of RLO, which
is the sequence with the lowest initial donor mass that manages to
satisfy all observational constraints for $GRS\, 1915+105$. The dark
grey (red) line corresponds to a MT sequence with
$M_2=10.0\rm\,M_{\odot}$, $M_{\rm BH}=7.0\,\rm M_{\odot}$, and $P_{\rm
  orb}=0.85\,\rm days$ at the start of RLO, which is the MT sequence
with the most massive donor. Figure~\ref{MTtracks} shows for the same
two sequences the evolution of the MT rate as a function of the time
elapsed since the onset of RLO.

\begin{figure}
\centering
\includegraphics[width=0.5\textwidth]{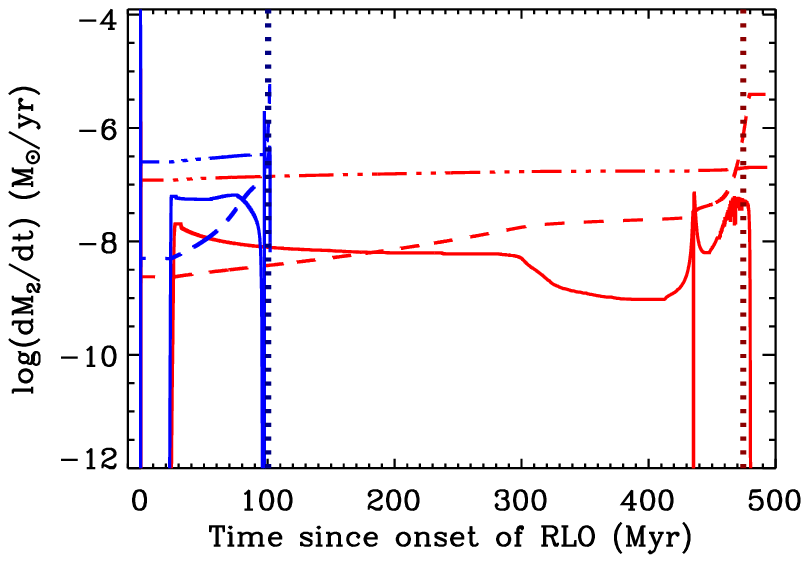}
\caption{\label{MTrate}The MT rate as a function of time since the onset of RLO for the same two MT sequences considered in Figure~4. The triple-dot-dashed lines correspond to the Eddington MT rate, and the dashed line to the critical MT rate below which the accretion disk becomes thermally unstable and the XRB exhibits transient behavior. The thick, vertical dotted lines show the time at which each of the MT sequences satisfy simultaneously all the observational constraint of GRS\,1915+105.}
\end{figure}

We see that in both of these examples for GRS\,1915+105 that the initial mass of the donor star is higher than the initial mass of the BH. Conservative MT from the more massive component of a binary to the less massive one causes the orbit to initially shrink, until the mass ratio of the binary inverts (see right panel of Figure~\ref{MTtracks}). During this initial phase of MT the donor star is still on the main sequence. At the same time, however, the shrinkage of the orbit due to angular momentum conservation gives rise to high MT rates ($\sim 10^{-5}\,\rm M_{\odot}\, yr^{-1}$) which also brings the donor star out of thermal equilibrium. This is more obvious in the sequence with the more massive donor star, as $\sim1.5\,\rm M_{\odot}$ must be transferred to the BH before the mass ratio is inverted. Soon after the mass ratio of the binary is inverted, the orbit starts expanding as a result again of angular momentum conservation, and the binary goes into a long-lived phase of thermally stable MT, at much lower MT rates ($\sim 10^{-8}\,\rm M_{\odot}\, yr^{-1}$).

At some point the donor star leaves the main sequence and evolves
towards the giant branch. This brings the binary into a second phase of
thermally unstable MT, during which the whole envelope of the donor star
will be transferred onto the BH. The end of this last MT phase will
leave behind the naked helium core of the donor star, detached and
orbiting around the BH. The low-mass naked helium core will become
degenerate and form a helium white dwarf. (See in the left panel of
Figure~\ref{MTtracks} the spike in the effective temperature of the
companion star toward the end of the MT sequence.) The evolutionary path
described here is typical for systems like GRS\,1915+105 where the BH
needs to double its mass via accretion in order to spin up all the way
to $a_*\sim 1$ \citep[see also discussion in][]{PRH2003}. Furthermore,
in this evolutionary scenario for GRS\,1915+105, a short phase of
super-Eddington accretion ($2-100\times \dot{M}_{\rm Edd}$) is
predicted, during which $0.5-6.4\rm M_\odot$ are accreted onto the BH
(see also Tables \ref{sucMT_noSpin} and \ref{sucMT_Spin}). Our results
are in agreement with \citet{PRH2003}, who first demonstrated that the
secondary of GRS\,1915+105 could have been an intermediate-mass
star. Their calculations, which assumed Eddington limited accretion,
showed that the initial mass of the donor in GRS\,1915+105 may have been
as high as $\sim 6\,\rm M_{\odot}$ and the BH may have accreted up to
$\sim 4\,\rm M_{\odot}$ from its companion. We should note here that
\citet{PRH2003} only considered an initial BH mass of $10\,\rm
M_{\odot}$, which is relatively massive and makes the spin-up of the BH
more difficult, and results in somewhat different observational
constraints for the current mass of the BH and its companion from
earlier observational studies. Taking into account these two factors, we
find that our constraints on the properties of GRS\,1915+105 at the
onset of RLO are in overall agreement with those from \citet{PRH2003},
and that our assumption of fully conservative (instead of Eddington
limited) accretion does not change the qualitative picture of the
evolution of this system nor the fact that the observed extreme spin can
be accounted for solely by the material that the BH has accreted after
its formation.

Figures \ref{MTtracks} and \ref{MTrate} showcase the evolution of the XRB with the most extreme measured BH spin. We stress that the evolution of all other systems with measured or estimated BH spins listed in Table \ref{ObsProp} does not require highly super-Eddington MT phases. Apart from GRS\,1915+105, the rest of the systems listed in Table \ref{ObsProp} can be divided into two categories. In the first category we have LMXBs in close orbits ($P_{\rm orb} < 1\rm\,day$) with K-dwarf donor stars. These systems followed an evolution similar to XTE J1118+480 \citep{Fragos2009b}, where the orbital period at the onset of RLO was below the bifurcation period ($\sim0.7-1.0\rm\, day$) and the initial donor mass was below $~1.5\rm\, M_\odot$, which allows magnetic braking to operate. The MT rate in these systems is regulated by the angular momentum losses due to magnetic braking, and is always well below the Eddington limit. In these systems a maximum of $\sim 1\,\rm M_\odot$ of material can be transferred from the donor star to the BH. Hence, the BHs in this subclass of LMXBs cannot be spun up via accretion, and their spins are therefore expected to be low.

In the second category we have XRBs with orbital periods of $\sim 1-3\,\rm days$ and, at the end of their main-sequence or subgiant phase, have donor stars of up to a few solar masses. The general evolutionary history of these systems is similar to that of GRO 1655-40, which was studied in detail by \citet{BP2002} and \citet{Willems2005}. The MT rate of this subclass of LMXBs is determined by the nuclear evolution of the donor which is still near the end of the main sequence and is expected to be again sub-Eddington \citep[see also][]{PRH2003}. The initial donor mass of this sub-class of LMXBs at the onset of RLO could be as high as $\sim 5\,\rm M_\odot$, providing a larger supply of material that can be accreted by the BH, resulting in a higher BH spin. Hence, the BHs in these systems can have moderately high spins.

\begin{deluxetable*}{ccccccccccccc}
\tablecolumns{13}
\tabletypesize{\scriptsize}
\tablecaption{Summary of selected properties of MT sequences calculated to satisfy simultaneously all the observational constraints, excluding the BH spin, for each of the BH LMXBs in Table~\ref{ObsProp}. The current parameters correspond to the point where the binary's orbital period is equal to the observed orbital period of each system.    
\label{sucMT_noSpin}}
\tablehead{ 
   \colhead{} & 
   \multicolumn{5}{c}{Parameters at onset of RLO} &
   \colhead{} &
   \multicolumn{6}{c}{Current parameters}  \\
   \cline{2-6} \cline{8-13}  \\
   \colhead{System} & 
   \colhead{$M_{\rm BH}$\tablenotemark{a}} & 
   \colhead{$M_2$\tablenotemark{b}} & 
   \colhead{$P_{\rm orb}$\tablenotemark{c}} &
   \colhead{$X_2$\tablenotemark{d}} &
   \colhead{$\tau_2$\tablenotemark{e}} &
   \colhead{} &
   \colhead{$M_{\rm BH}$\tablenotemark{a}} & 
   \colhead{$M_2$\tablenotemark{b}} & 
   \colhead{$X_2$\tablenotemark{d}} &
   \colhead{$\tau_2$\tablenotemark{e}} &
   \colhead{$max.\,M_{\rm acc}$\tablenotemark{g}} &
   \colhead{$max.\,a_*$\tablenotemark{h}} \\
   \colhead{} & 
   \colhead{($M_\odot$)} & 
   \colhead{($M_\odot$)} & 
   \colhead{(days)} &
   \colhead{} &
   \colhead{(Gyr)} &
   \colhead{} &
   \colhead{($M_\odot$)} & 
   \colhead{($M_\odot$)} & 
   \colhead{} &
   \colhead{(Gyr)} &
   \colhead{($M_\odot$)} &
   \colhead{} 
   }
\startdata
\multicolumn{12}{c}{} \\
GRS 1915+105                        &  3.-10. & 1.0-10.0& 0.6-30.0 & 0.0-0.7 &  0.0-12.4 & &  8.4-15.9 & 0.3-1.3 & 0.0-0.0 & 0.1-12.5 & 0.0-9.0\,\it{(0.0-5.6)} & 0.01-1.00\,\emph{(0.01-0.95)} \\
4U 1543-47                          &  3.-10. & 2.2-6.4 & 0.6- 1.1 & 0.3-0.7 &  0.0- 0.5 & &  5.4-10.8 & 2.0-2.7 & 0.3-0.4 & 0.1- 0.6 & 0.0-4.0\,\it{(0.0-1.4)} & 0.01-1.00\,\emph{(0.00-0.76)} \\
GRO J1655-40$^{GBO}$                &  4.- 6. & 2.6-5.0 & 0.7- 1.7 & 0.2-0.7 &  0.0- 0.4 & &  5.4- 7.3 & 1.7-2.1 & 0.0-0.2 & 0.3- 0.6 & 0.5-3.2\,\it{(0.5-2.7)} & 0.26-0.94\,\emph{(0.26-0.90)} \\
GRO J1655-40$^{BP}$                 &  3.- 5. & 2.6-3.8 & 1.0- 1.7 & 0.2-0.5 &  0.1- 0.4 & &  4.8- 5.9 & 1.7-2.0 & 0.0-0.2 & 0.2- 0.6 & 0.6-2.1\,\it{(0.5-1.8)} & 0.33-0.90\,\emph{(0.33-0.78)} \\
XTE J1550-564                       &  7.- 9. & 0.9-1.5 & 0.3- 0.9 & 0.0-0.6 &  1.1-10.6 & &  8.0-10.0 & 0.2-0.4 & 0.0-0.6 & 4.0-13.5 & 0.6-1.2\,\it{(0.6-1.2)} & 0.21-0.44\,\emph{(0.21-0.44)} \\
A0620-00                            &  5.- 6. & 1.1-1.8 & 0.6- 0.8 & 0.0-0.7 &  0.0- 6.4 & &  6.3- 6.9 & 0.4-0.5 & 0.0-0.1 & 2.7- 8.0 & 0.7-1.3\,\it{(0.7-1.3)} & 0.34-0.59\,\emph{(0.34-0.59)} \\
GRS 1124-683                        &  4.- 8. & 1.0-1.8 & 0.3- 0.9 & 0.0-0.7 &  0.0-10.2 & &  4.8- 8.9 & 0.3-0.8 & 0.0-0.6 & 0.9-11.9 & 0.3-1.1\,\it{(0.3-1.1)} & 0.12-0.62\,\emph{(0.12-0.62)} \\
GX 339-4                            &  3.- 9. & 0.6-8.8 & 0.2- 1.7 & 0.0-0.7 &  0.0-11.9 & &  6.0-10.0 & 0.0-3.0 & 0.0-0.7 & 0.0-13.6 & 0.0-5.8\,\it{(0.0-2.6)} & 0.01-1.00\,\emph{(0.00-0.89)} \\
XTE J1859+226                       &  5.- 9. & 0.6-1.8 & 0.2- 0.9 & 0.0-0.7 &  0.0-10.2 & &  6.1-10.0 & 0.0-1.1 & 0.0-0.7 & 0.1-13.6 & 0.1-1.5\,\it{(0.0-1.5)} & 0.02-0.63\,\emph{(0.00-0.63)} \\
GS 2000+251                         &  5.- 9. & 0.9-1.8 & 0.3- 0.9 & 0.0-0.7 &  0.0-10.2 & &  6.0- 9.9 & 0.2-0.8 & 0.0-0.6 & 0.8-13.0 & 0.1-1.3\,\it{(0.1-1.3)} & 0.05-0.57\,\emph{(0.05-0.57)} \\
GRO J0422+32                        &  5.- 9. & 0.8-1.5 & 0.3- 0.7 & 0.2-0.7 &  0.0-10.2 & &  6.0-10.0 & 0.5-0.6 & 0.1-0.7 & 0.4-10.8 & 0.2-1.0\,\it{(0.0-1.0)} & 0.09-0.49\,\emph{(0.00-0.49)} \\
GRS 1009-45                         &  6.-10. & 1.0-1.6 & 0.6- 0.8 & 0.0-0.7 &  0.0- 9.2 & &  6.5-10.5 & 0.3-0.6 & 0.0-0.3 & 2.1-10.4 & 0.5-1.3\,\it{(0.5-1.3)} & 0.15-0.50\,\emph{(0.15-0.50)} \\
GS 1354-64                          &  3.- 9. & 1.6-6.8 & 0.6- 2.4 & 0.0-0.7 &  0.0- 2.1 & &  6.0-10.0 & 1.2-1.9 & 0.0-0.1 & 0.4- 2.1 & 0.0-5.1\,\it{(0.0-2.5)} & 0.01-1.00\,\emph{(0.01-0.89)} \\
GS 2023+338                         &  7.- 9. & 1.0-2.0 & 0.6- 2.0 & 0.0-0.7 &  0.0-12.0 & &  7.8-10.2 & 0.5-0.6 & 0.0-0.0 & 2.1-12.3 & 0.4-1.4\,\it{(0.4-1.4)} & 0.13-0.49\,\emph{(0.13-0.49)} \\
H1705-250                           &  4.- 6. & 1.0-1.5 & 0.4- 0.9 & 0.0-0.7 &  0.0- 5.4 & &  5.2- 7.4 & 0.1-0.3 & 0.0-0.5 & 2.3- 6.6 & 0.9-1.4\,\it{(0.9-1.4)} & 0.40-0.63\,\emph{(0.40-0.63)} \\
V4641 Sgr\tablenotemark{$\dagger$}  &  3.- 4. & 7.0-7.8 & 1.2- 1.7 & 0.3-0.5 &  0.0- 0.0 & &  5.4- 6.6 & 2.1-2.7 & 0.0-0.2 & 0.0- 0.1 & 2.3-2.6\,\it{(0.0-0.9)} & 0.85-0.94\,\emph{(0.03-0.53)} \\
XTE J1118+480                       &  6.- 7. & 1.0-1.8 & 0.6- 0.8 & 0.0-0.7 &  0.0- 9.2 & &  6.9- 8.1 & 0.1-0.3 & 0.0-0.6 & 1.5-11.2 & 0.7-1.6\,\it{(0.7-1.6)} & 0.29-0.59\,\emph{(0.29-0.59)} \\

\enddata
\tablenotetext{a}{BH mass}
\tablenotetext{b}{Donor star mass}
\tablenotetext{c}{Orbital period}
\tablenotetext{d}{Central hydrogen fraction of the donor star}
\tablenotetext{e}{Age of the donor star}
\tablenotetext{f}{Donor's effective temperature}
\tablenotetext{g}{Maximum amount of mass that the BH can accrete, assuming fully conservative MT. The numbers in parenthesis is the amount of mass that would have been accreted assuming that the accretion rate could not exceed the Eddington limit.} 
\tablenotetext{h}{Maximum spin parameter $a_*$ that the BH can acquire, assuming fully conservative MT. The numbers in parenthesis is the spin parameter that the BH would have if the accretion rate could not exceed the Eddington limit.} 
\tablenotetext{$\dagger$}{No ``succesful'' MT sequences were found for V4641\,Sgr assuming fully conservative MT. The values reported for this system are assuming an accretion efficiency of 50\%. See text for details.}
\end{deluxetable*}

Following our methodology, we are able to find ``succesful'' MT
sequences for \emph{all} of the Galactic BH LMXBs listed in
Table~\ref{ObsProp}. As explained earlier, these MT sequences represent
possible progenitors of the currently observed
systems. Table~\ref{sucMT_noSpin} shows a summary of the selected
properties of MT sequences calculated to satisfy simultaneously all the
observational constraints, excluding the BH spin measurement or
estimate, for each of the BH LMXBs in Table~\ref{ObsProp}. The current
parameters correspond to the point where for each system the binary's
orbital period is equal to the observed orbital period. As mentioned
earlier, we assumed fully conservative MT, and thus the estimated amount
of accreted material and the estimated BH spin should only be thought of
as upper limits. This is why the last two columns of
Tables~\ref{sucMT_noSpin}-\ref{sucMT_All} are denoted as the maximum
accreted mass ($Max.~M_{\rm acc}$) and maximum BH spin ($Max.~a_*$). In
addition, we report for comparison (in parentheses) the amount of mass
that would have been accreted and the spin parameter that the BH would
have achieved assuming that the accretion rate could not exceed the
Eddington limit. The numbers reported in parentheses have not been
calculated self-consistently; they are simply estimates derived by
calculating for each sequence how much of the transferred material was
accreted at MT rates exceeding the Eddington limit.

For V4641\,Sgr we were only able to find ``successful'' MT sequences
that satisfy all the observational constraints simultaneously by
assuming an accretion efficiency of 50\% and a MT that is not fully
conservative. The companion star in V4641\,Sgr is too hot for a star of
its mass, suggesting that it has a larger core than an isolated star of
$\sim 3\,\rm M_{\odot}$ at a similar evolutionary stage.  Such a
condition in an interacting binary may result from the donor star losing
part of its envelope while its core remains approximately its original
size, thus appearing hotter and more luminous than indicated by its
current mass. At the same time, however, if the BH accreted all the
material the companion star lost in order to appear as it is today, then
the BH mass would exceed its currently observed value. This issue, which
arises only in the case of V4641\,Sgr, results from the properties of
the donor star described above and the precise value of mass reported
for this BH. Furthermore, this is the only system in our study for which
we find that the companion was most likely significantly more massive
than the BH at the onset of the RLO. This results in an initial phase of
thermally unstable MT phase, with MT rates as high as $\sim 10^{-3}$,
which persists until the mass ratio of the binary inverts. Our
assumption of fully conservative MT, given these extreme super-Eddington
MT rates, is highly unlikely in this case. This in turn justifies our
finding that an accretion efficiency of $\lesssim50\%$ is needed in
order to explain the evolutionary history of V4641\,Sgr.

Our grid of MT sequences was calculated assuming fully conservative
MT. The estimate of the effect of a lower accretion efficiency (similar
to the situation described earlier concerning the MT rate pegged at the
Eddington limit) was done at post processing in an approximate
way. Namely, for a MT sequence that was calculated initially as fully
conservative, we simply recalculated the BH mass at each timestep
assuming that only part of the material lost from the donor was accreted
by the BH. There is an inconsistency in this approach, as the orbital
evolution of the binary is still done assuming fully conservative
MT. However, this approximate estimate already gives us a good feeling
for the effects of an accretion efficiency that is less than 100\%.  A
proper determination of these numbers would require a recalculation of
the entire grid of MT sequences in order to self-consistently evolve the
orbit of the binary when part of the material is lost from the
system. However, this would be very computationally expensive and is
outside the scope of this paper.

\subsection{Constraints on the Maximum BH Spin}

Following the same procedure as described in Section 5.1, but this time
using as an additional constraint, namely the measured or estimated BH
spin for those systems for which these data are available, we can test
the main hypothesis of this work: That BHs in Galactic LMXBs were born
with negligible spin, and that their currently observed spin is an
effect of mass accretion after BH formation. Table~\ref{sucMT_Spin}
shows that for each of the 9 Galactic BH LMXBs from Table~\ref{ObsProp}
that have a BH spin measurement or estimate, we are able to find
``succesful'' MT sequences that satisfy simultaneously all the
observational constraints, \emph{including} the BH's spin. Therefore,
our principal hypothesis is viable. Of course, the MT sequences
summarized for each observed system in Table~\ref{sucMT_Spin} are a
subset of the MT sequences summarized in Table~\ref{sucMT_noSpin}. A
detailed list of each ``successful'' MT sequence and its properties at
the onset of RLO and its properties presently, with and without the
observed BH constraint, can be found in Table~\ref{sucMT_All}.

\begin{deluxetable*}{ccccccccccccc}
\tablecolumns{13}
\tabletypesize{\scriptsize}
\tablecaption{Same as Table~\ref{sucMT_noSpin}, but this time the MT
  sequences  satisfy simultaneously all the observational constraints,
  including the BH spin measurement or estimate. Only the 9 BH LMXBs for
  which a BH spin measurement or estimate exists are listed in this table.
\label{sucMT_Spin}}
\tablehead{ 
\colhead{} & 
\multicolumn{5}{c}{Parameters at onset of RLO} &
\colhead{} &
\multicolumn{6}{c}{Current parameters}  \\
\cline{2-6} \cline{8-13}  \\
\colhead{System} & 
\colhead{$M_{\rm BH}$\tablenotemark{a}} & 
\colhead{$M_2$\tablenotemark{b}} & 
\colhead{$P_{\rm orb}$\tablenotemark{c}} &
\colhead{$X_2$\tablenotemark{d}} &
\colhead{$\tau_2$\tablenotemark{e}} &
\colhead{} &
\colhead{$M_{\rm BH}$\tablenotemark{a}} & 
\colhead{$M_2$\tablenotemark{b}} & 
\colhead{$X_2$\tablenotemark{d}} &
\colhead{$\tau_2$\tablenotemark{e}} &
\colhead{$max.\,M_{\rm acc}$\tablenotemark{g}} &
\colhead{$max.\,a_*$\tablenotemark{h}} \\
\colhead{} & 
\colhead{($M_\odot$)} & 
\colhead{($M_\odot$)} & 
\colhead{(days)} &
\colhead{} &
\colhead{(Gyr)} &
\colhead{} &
\colhead{($M_\odot$)} & 
\colhead{($M_\odot$)} & 
\colhead{} &
\colhead{(Gyr)} &
\colhead{($M_\odot$)} &
\colhead{} 
   }
\startdata
\multicolumn{12}{c}{} \\
GRS 1915+105        &  3.- 7. & 5.2-10.0 & 0.7- 1.7 & 0.3-0.7 &  0.0- 0.1 & &  8.4-15.7 & 0.6-1.3 & 0.0-0.0 & 0.1- 0.9 & 4.5-9.0\,\it{(0.5-5.1)} & 0.98-1.00\,\emph{(0.36-0.94)} \\
4U 1543-47          &  3.- 4. & 3.8-6.4 & 0.6- 0.8 & 0.6-0.7 &  0.0- 0.0 & &  5.4- 7.0 & 2.1-2.5 & 0.3-0.4 & 0.1- 0.5 & 1.4-4.0\,\it{(0.1-1.4)} & 0.69-1.00\,\emph{(0.13-0.76)} \\
GRO J1655-40$^{GBO}$&  4.- 6. & 3.0-5.0 & 0.7- 1.3 & 0.3-0.7 &  0.0- 0.2 & &  5.4- 7.3 & 1.7-2.0 & 0.1-0.2 & 0.3- 0.4 & 1.0-3.2\,\it{(1.0-2.7)} & 0.50-0.94\,\emph{(0.50-0.90)} \\
GRO J1655-40$^{BP}$&  3.- 4. & 2.6-3.8 & 1.0- 1.5 & 0.2-0.5 &  0.1- 0.4 & &  4.8- 5.8 & 1.7-1.9 & 0.0-0.2 & 0.2- 0.6 & 0.9-2.1\,\it{(0.5-1.8)} & 0.52-0.90\,\emph{(0.43-0.78)} \\
XTE J1550-564       &  7.- 9. & 0.9-1.5 & 0.3- 0.9 & 0.0-0.6 &  1.1-10.6 & &  8.0-10.0 & 0.2-0.4 & 0.0-0.6 & 4.0-13.5 & 0.6-1.2\,\it{(0.6-1.2)} & 0.21-0.44\,\emph{(0.21-0.44)} \\
A0620-00            &  6.- 6. & 1.1-1.3 & 0.7- 0.8 & 0.0-0.1 &  2.8- 6.4 & &  6.7- 6.9 & 0.4-0.5 & 0.0-0.0 & 4.1- 8.0 & 0.7-0.9\,\it{(0.7-0.9)} & 0.34-0.39\,\emph{(0.34-0.39)} \\
GRS 1124-683        &  4.- 8. & 1.0-1.8 & 0.3- 0.9 & 0.0-0.7 &  0.0-10.2 & &  4.8- 8.9 & 0.3-0.8 & 0.0-0.6 & 0.9-11.9 & 0.3-1.1\,\it{(0.3-1.1)} & 0.12-0.47\,\emph{(0.12-0.47)} \\
GX 339-4            &  5.- 9. & 0.6-4.4 & 0.2- 1.7 & 0.0-0.7 &  0.0-11.9 & &  6.0-10.0 & 0.0-2.6 & 0.0-0.7 & 0.1-13.6 & 0.0-1.8\,\it{(0.0-1.6)} & 0.01-0.55\,\emph{(0.00-0.55)} \\
XTE J1859+226       &  5.- 9. & 0.6-1.8 & 0.2- 0.9 & 0.0-0.7 &  0.0-10.2 & &  6.1-10.0 & 0.0-1.1 & 0.0-0.7 & 0.1-13.6 & 0.1-1.2\,\it{(0.0-1.2)} & 0.02-0.54\,\emph{(0.00-0.54)} \\
GS 2000+251         &  6.- 9. & 0.9-1.1 & 0.3- 0.6 & 0.0-0.7 &  0.7-10.2 & &  6.2- 9.4 & 0.6-0.8 & 0.0-0.6 & 0.8-10.3 & 0.1-0.4\,\it{(0.1-0.4)} & 0.05-0.14\,\emph{(0.05-0.14)} \\

\enddata
\tablenotetext{a}{BH mass}
\tablenotetext{b}{Donor star mass}
\tablenotetext{c}{Orbital period}
\tablenotetext{d}{Central hydrogen fraction of the donor star}
\tablenotetext{e}{Age of the donor star}
\tablenotetext{f}{Donor's effective temperature}
\tablenotetext{g}{Maximum amount of mass that the BH can accrete assuming fully conservative MT. The numbers in parentheses are the amount of mass that would have been accreted assuming that the accretion rate could not exceed the Eddington limit.} 
\tablenotetext{h}{Maximum spin parameter $a_*$ that the BH can acquire assuming fully conservative MT. The numbers in parentheses are the spin parameter that the BH would have if the accretion rate could not exceed the Eddington limit.} 
\end{deluxetable*}

\begin{deluxetable*}{lcccccccccccccc}
\tablecolumns{15}
\tabletypesize{\scriptsize}
\tablecaption{Selected properties of all individual MT sequences calculated to satisfy simultaneously all the observational constraints for each of the BH LMXBs in Table~\ref{ObsProp}. The current parameters correspond to the point where the binary's orbital period is equal to the observed orbital period of each system. Boldface denotes MT sequences that satisfy simultaneously all the observational constraints including the measured or estimated BH spin whenever it is available.
\label{sucMT_All}}
\tablehead{ 
\colhead{} & 
\multicolumn{5}{c}{Parameters at onset of RLO} &
\colhead{} &
\multicolumn{7}{c}{Current parameters}  \\
\cline{2-7} \cline{9-15}  \\
\colhead{System} & 
\colhead{Sequence} & 
\colhead{$M_{\rm BH}$\tablenotemark{a}} & 
\colhead{$M_2$\tablenotemark{b}} & 
\colhead{$P_{\rm orb}$\tablenotemark{c}} &
\colhead{$X_2$\tablenotemark{d}} &
\colhead{$\tau_2$\tablenotemark{e}} &
\colhead{} &
\colhead{$M_{\rm BH}$\tablenotemark{a}} & 
\colhead{$M_2$\tablenotemark{b}} & 
\colhead{$log(T_{\rm eff})$\tablenotemark{f}} & 
\colhead{$X_2$\tablenotemark{d}} &
\colhead{$\tau_2$\tablenotemark{e}} &
\colhead{$max.\,M_{\rm acc}$\tablenotemark{g}} &
\colhead{$max.\,a_*$\tablenotemark{h}} \\
\colhead{} & 
\colhead{} & 
\colhead{($M_\odot$)} & 
\colhead{($M_\odot$)} & 
\colhead{(days)} &
\colhead{} &
\colhead{(Gyr)} &
\colhead{} &
\colhead{($M_\odot$)} & 
\colhead{($M_\odot$)} & 
\colhead{($K$)} & 
\colhead{} &
\colhead{(Gyr)} &
\colhead{($M_\odot$)} &
\colhead{} 
   }
\startdata
\multicolumn{15}{c}{} \\
GRS 1915+105& & & & & & & & & & & & & & \\
 &       1 &  8.0 & 1.0 &  1.30 & 0.00 & 11.76 & &  8.7 & 0.29 & 3.63 & 0.00 & 12.48 & 0.71\,\it{(0.70)} & 0.26\,\emph{(0.26)} \\
 &       2 &  8.0 & 1.0 &  1.40 & 0.00 & 11.83 & &  8.7 & 0.30 & 3.62 & 0.00 & 12.46 & 0.70\,\it{(0.69)} & 0.26\,\emph{(0.26)} \\
  & $\vdots$ & $\vdots$ & $\vdots$ & $\vdots$ & $\vdots$ & $\vdots$ & $\vdots$ & $\vdots$ & $\vdots$ & $\vdots$ & $\vdots$ & $\vdots$ & $\vdots$ & $\vdots$ \\
 &    1854 & 10.0 & 5.2 &  1.50 & 0.32 &  0.06 & & 13.9 & 1.33 & 3.69 & 0.00 &  0.11 & 3.87\,\it{(3.71)} & 0.73\,\emph{(0.71)} \\
 & \bf   1855 &  4.0 & 5.4 &  0.70 & 0.70 &  0.00 & &  8.7 & 0.67 & 3.63 & 0.00 &  0.47 & 4.73\,\it{(3.05)} & 0.99\,\emph{(0.93)} \\
 &    1856 &  5.0 & 5.4 &  1.20 & 0.45 &  0.04 & &  9.5 & 0.94 & 3.67 & 0.00 &  0.16 & 4.46\,\it{(3.35)} & 0.96\,\emph{(0.90)} \\
  & $\vdots$ & $\vdots$ & $\vdots$ & $\vdots$ & $\vdots$ & $\vdots$ & $\vdots$ & $\vdots$ & $\vdots$ & $\vdots$ & $\vdots$ & $\vdots$ & $\vdots$ & $\vdots$ \\
4U 1543-47& & & & & & & & & & & & & & \\
 &       1 &  6.0 & 2.2 &  1.00 & 0.37 &  0.50 & &  6.1 & 2.07 & 3.91 & 0.28 &  0.60 & 0.13\,\it{(0.12)} & 0.07\,\emph{(0.07)} \\
 &       2 &  6.0 & 2.2 &  1.10 & 0.33 &  0.55 & &  6.0 & 2.18 & 3.92 & 0.27 &  0.61 & 0.02\,\it{(0.01)} & 0.01\,\emph{(0.01)} \\
 $\vdots$ & $\vdots$ & $\vdots$ & $\vdots$ & $\vdots$ & $\vdots$ & $\vdots$ & $\vdots$ & $\vdots$ & $\vdots$ & $\vdots$ & $\vdots$ & $\vdots$ & $\vdots$ & $\vdots$ \\

\enddata
\tablenotetext{a}{BH mass}
\tablenotetext{b}{Donor star mass}
\tablenotetext{c}{Orbital period}
\tablenotetext{d}{Central hydrogen fraction of the donor star}
\tablenotetext{e}{Age of the donor star}
\tablenotetext{f}{Donor's effective temperature}
\tablenotetext{g}{Maximum amount of mass that the BH can accrete assuming fully conservative MT. The numbers in parentheses are the amount of mass that would have been accreted assuming that the accretion rate could not exceed the Eddington limit.} 
\tablenotetext{h}{Maximum spin parameter $a_*$ that the BH can acquire assuming fully conservative MT. The numbers in parentheses are the spin parameter that the BH would have if the accretion rate could not exceed the Eddington limit.} 

\end{deluxetable*}

Having proven the viability of our hypothesis, we can now use our grid of MT sequences and current observational data to make predictions about the maximum spin of the BH in those systems for which there is presently no measurement or estimate of spin. These predictions are listed in Table~\ref{sucMT_noSpin}. As summarized in Table~\ref{ObsProp}, we predict for 7 out of the 11 LMXBs without spin measurements or estimates, a spin parameter of $\lesssim 0.6$. This is a strong prediction that will be verified or falsified in the next few years as more BH spin measurements become available. The accuracy of our predicted maximum spins depends for each BH always on the accuracy with which the other observational properties of the LMXB, namely BH and donor mass, effective temperature of the donor star and orbital period, are determined. As the quality of the observational constraints improve over time, so will the accuracy of the limits we set on the BH spin.

\begin{figure*}
\centering
\includegraphics[width=1.0\textwidth]{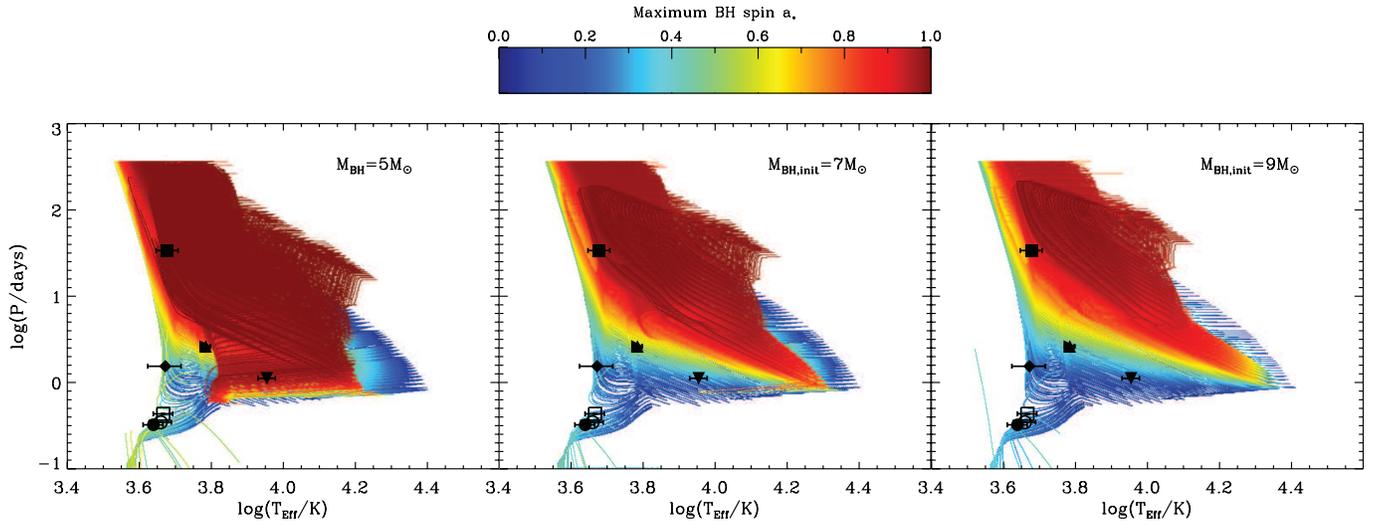}
\caption{The colormaps show the \emph{maximum} BH spin that a Galactic
  LMXB can have based on the hypothesis that BHs in LMXBs acquire their
  spin through accretion, for a given orbital period and effective
  temperature of the donor star and for three different birth BH
  masses. For comparison we overplot the orbital period and donor star's
  effective temperature for the 7 systems in Table~\ref{ObsProp},
  for which an estimate of the effective temperature is
  available. \label{SpinMap}}
\end{figure*}

Apart from predicting the maximum BH spin of a specific Galactic BH
LMXB, our analysis also makes predictions about a general BH LMXB
population. In principle, based on our grid of MT calculations, for any
four values of the observable properties of BH and donor star mass,
effective temperature of the donor star and orbital period, we can first
identify whether a LMXB with this combination of properties can exist;
if so, then, subsequently, if the combination of 4 observable properties
is covered by our grid, we can estimate the maximum spin that a BH can
have in a system with such properties. However, there is no easy and
intuitive way to visualize such a multidimensional parameter
space. Figure~\ref{SpinMap} shows slices of this four-dimensional
parameter space along the orbital period and donor star's effective
temperature, for three different values of initial BH mass. The
colormaps show the maximum BH spin that a Galactic LMXB can have, based
on the hypothesis that BHs in LMXBs acquire their spin through
accretion. Of course, some information has been lost in the process of
creating this colormap, as we have marginalized over the donor-star mass
axis. We again emphasize that the colormap of Figure~\ref{SpinMap} shows
the \emph{maximum} spin that a BH can have, based on other properties of
the binary it resides in. The reason that we can only set an upper limit
on the BH spin (and are not able to estimate an actual value) is due to
the poor constraints that we have, both observationally and
theoretically, on the accretion efficiency. Our grid of MT calculations
considers the limiting case of fully conservative MT (i.e. an accretion
efficiency of 100\%); hence our predictions of BH spin should only be
treated as upper limits.

\begin{figure}
\centering
\includegraphics[width=0.5\textwidth]{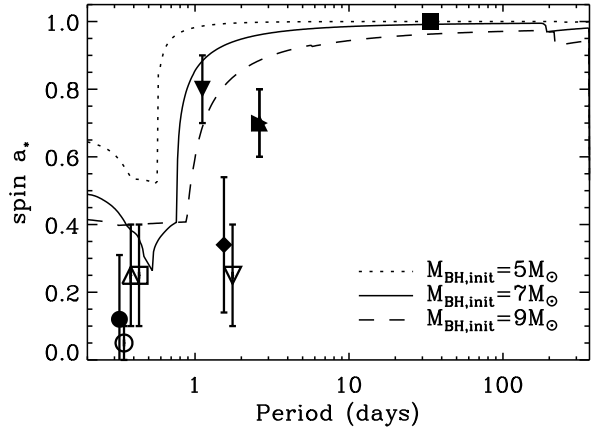}
\caption{The \emph{maximum} BH spin that a Galactic LMXB can have based
  on the hypothesis that BHs in LMXBs acquire their spin through
  accretion, for a given orbital period, for three different birth BH
  masses. For comparison we over-plot the measured or estimated BH spin
  versus the orbital period for the 9 LMXBs systems in
  Table~\ref{ObsProp}. All observed systems appear to be either close or
  below the predicted lines of maximum BH spin.\label{SpinOrb}}
\end{figure}

In all three panels of Figure~\ref{SpinMap}, there is a region of parameter space, for orbital periods either close or below the bifurcation period ($\lesssim 18\,\rm hr$), where our analysis predicts that the angular momentum a BH can gain through accretion is quite limited and the maximum spin that it can achieve is low ($a_*\lesssim 0.5$). This result becomes more evident if we marginalize the colormaps of Figure~\ref{SpinMap} along the effective temperature axis. Figure~\ref{SpinOrb} shows exactly this. The three lines show the maximum BH spin that a Galactic LMXB can have at a given orbital period for three different birth BH masses. Although these curves offer very weak constraints for orbital periods $\gtrsim 1\,\rm day$, they pose strong constraints for the spin of BHs that reside in tight binaries.

The qualitative physical explanation of this result is simple. Due to angular momentum conservation, MT from the less massive component of a binary to the more massive one, as is the case in BH LMXBs, results in orbital expansion. Thus, after the onset of RLO in a BH LMXB, the orbital period should only increase. The mass-radius relation of zero-age main-sequence stars tells us that companion stars that fit in tight orbits of several hours have main-sequence lifetimes longer than the Hubble time. Hence, such a star would not have enough time to expand via nuclear evolution, fill its Roche lobe and enter the MT phase. However, the majority of the observed BH LMXBs in the Milky Way are found in binaries with orbital periods less than 18 hours, all the way down to a few hours. These systems must have initiated the MT in a wider binary with a more massive companion star, and due to some angular momentum loss mechanism the orbit shrank instead of expanding, as one would expect from angular momentum conservation arguments. This additional angular momentum loss mechanism is magnetic braking, and it is known to operate in all low-mass stars ($\lesssim1.5\,\rm M_{\odot}$) which have a convective envelope and a radiative core. Since any BH LMXB observed today at a period below the bifurcation period must have had magnetic braking operating in order to shrink the orbit to its currently observed period, the initial mass of the companion star must have been $\lesssim1.5\,\rm M_{\odot}$. The upper limit on the initial mass of the donor star puts also an upper limit on the mass that the BH can have accreted during the XRB phase. BHs found in tight LMXBs can only have accreted up to $\sim 1\,\rm M_{\odot}$ of material from the donor star, and therefore they can only be mildly spun up. Figure~\ref{SpinOrb} clearly demonstrates this effect.

\subsubsection{The reported retrograde BH spin in GRS 1124-683}

\citet{Morningstar2014} recently reported a retrograde spin for
GRS\,1124-683 ($a_* = -0.25_{-0.64}^{+0.05}$ at 90\% confidence) based
on a continuum-fitting analysis, which is a surprising result. As noted
in Section 1.1, this result is quite uncertain, and we do not make use
of it in this paper. We nevertheless now discuss the implications of a
retrograde spin.

There are three possible formation scenarios that can lead to a
retrograde BH spin. One way to create such a system is with a large
supernova kick, which is finally tuned in direction, imparted to the BH
during its formation. However, apart from the fact that only a very
small part of the parameter space would lead to such a system, making it
very rare, one would also expect to see this system currently flying
through the galaxy with a very high systemic velocity. There is some
evidence that GRS\,1124-683 may have received a non-negligible kick
\citep[$\gtrsim 65\,\rm km\, s^{-1}$;][see Table 5]{Repetto2012}, a kick
magnitude though, that does not make GRS\,1124-683 special compared to
other Galactic BH LMXBs. Alternatively, there could have been an
off-center asymmetry in the supernova explosion that changed completely
both the spin axis and the spin magnitude, probably in a random way and
direction. However, we have no evidence that this is the usual case, as
the statistics of spin measurements so far do not point towards a random
distribution in either direction nor magnitude \citep[albeit
see][]{Farr2011b}. Finally, the eccentric Kozai-Lidov mechanism
\citep{Naoz2011, Naoz2013} has been shown to be a very effective in
producing close binaries in hierarchical triple systems, where
spin-orbit misalignment angle of the inner close binary can have values
all the way up to $180^{\circ}$ \citep{Naoz2014}. Hence, a hierarchical
triple origin of GRS\,1124-683 could in principle explain a retrograde
BH spin. The common caveat of all aforementioned possible formation
channels for GRS\,1124-683 is that, although they can produce a
misalignment of $\gtrsim 90^{\circ}$, the probability to create a
misalignment of exactly $\gtrsim 180^{\circ}$ is very small, if not
negligible \citep{Fragos2010, Naoz2014}. Meanwhile, the continuum
fitting method for the measurement of BH spin is based on the assumption
that the BH spin axis is aligned with the orbital angular momentum axis,
allowing for either $\sim0^{\circ}$ or $\sim180^{\circ}$ spin-orbit
misalignment angles.

In conclusion, given the importance of an established example of retrograde spin to our understanding of BH formation and BH binary evolution, it is important to confirm or correct the \citet{Morningstar2014} result via a rigorous continuum-fitting analysis of the X-ray data and improved determinations of the crucial input parameters: BH mass, inclination and distance (Section 1).

\subsubsection{The origin of BH spin in HMXBs}

As one can see in Table~\ref{spin_table} and Figure~\ref{spin_plot}, all three HMXBs with measured spins are wind fed systems which have, nevertheless, small orbital periods. In fact, in all three systems the companion stars are almost filling their Roche lobes. Taking into account the short lifetimes of these systems and the fact that the BH can only accrete mass from its companion's stellar wind, the amount of accreted mass is negligible and therefore cannot explain the observationally inferred BH spins. Hence, the BH spin in these systems must be natal \citep{Gou2011}. It is possible that all three have always been in tight orbits, since the birth of the binary. This would have allowed the tides to operate on spinning up the core of the BH progenitor, and hence the spin in these BHs can be natal.

Massive HMXBs, like M33 X-7, can form in the galactic fields without going through a common envelope phase \citep[e.g.][]{Valsecchi2010,Wong2012,Wong2014}. Instead, in a primordial massive binary with a short orbital period of a few days, the more massive primary evolves faster than the secondary, growing in size to accommodate the energy produced at its center. Eventually it expands and begins MT onto the secondary through RLO. During the first few tens of thousands of years of MT, the orbital period decreases because the more massive primary is transferring mass to the less massive secondary. When the secondary accretes enough matter to become the more massive component, the orbit starts expanding \citep{Verbunt1993}. The primary transfers most of its H-rich envelope and becomes a Wolf-Rayet star, and the strong Wolf-Rayet wind ($\sim 2-3\times 10^{-5}\,\rm M_{\odot}\,yr^{-1}$) removes much of the remaining envelope, eventually interrupting the MT. Once the Wolf-Rayet wind sets in and the MT is interrupted, the wind blows away the remaining primary's envelope to expose the helium core. At this stage, the helium star and the main-sequence companion star are still in an orbit of a few days. Assuming a full tidal synchronization during the MT phase and solid body rotation, the resulting massive helium star has enough angular momentum when the binary detaches that even a highly spinning BH can be produced. However, both of these assumption can be violated.

Among the three high-mass XRBs with measured BH spin, Cyg\,X-1 is the
one with the most extreme measured spin value, and perhaps the one that
is the most difficult to explain. \citet{PRH2003} presented a
possible evolutionary scenario for Cyg\, X-1 which suggests that at
present the system is likely in a wind MT phase, following an earlier
Roche-lobe overflow phase. In this formation scenario, the initial mass
of the BH was $\sim12\,\rm M_{\odot}$, with the secondary of initial
mass $\sim25\,\rm M_{\odot}$ filling its Roche-lobe at a period of $\sim
7\,\rm day$. MT occurs initially on the thermal timescale. When the mass
ratio of the binary inverts, the secondary reestablishes thermal
equilibrium and becomes detached as the continuing wind mass loss of the
donor widens the orbit. At this point, the BH accretes material only
from the stellar wind of the secondary and the system reaches its
currently observed state. Eventually the secondary will expand again
after it leaves the main sequence and will fill its Roche lobe for a
second time. This scenario would also explain the extreme spin of the
BH, which can be attributed to accretion of material during the RLO
phase. However, an important prediction of this formation scenario is
that currently the secondary star should be less massive than the BH,
something that is in disagreement with the most recent mass estimates
for the two binary components of Cyg\, X-1
\citep{2011ApJ...742...84O}. More recently, \citet{Axelsson2011}
presented theoretical arguments showing how tidal locking of the binary,
before BH formation, is unlikely to explain the spin of the BH in the
HMXB Cyg\,X-1.  Before reaching a definite conclusion, however, one
needs to follow in detail the internal stellar rotation and the angular
momentum exchange between the two stars and the orbit, both during the
MT phases and the detached evolution of the binary.  This is, of course,
outside the scope of the current paper.  Our purpose here is to
highlight that the significantly different formation channels of
low-mass and high-mass XRBs may imply different origins for the spins of
the BHs they host.

\subsection{Implications for the Birth BH-Mass Distribution}

\begin{figure*}
\centering
\includegraphics[width=0.49\textwidth]{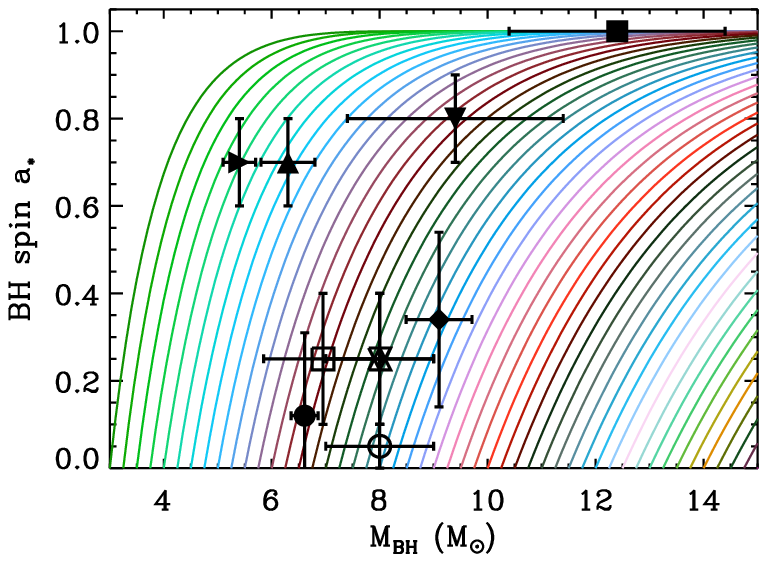}
\includegraphics[width=0.49\textwidth]{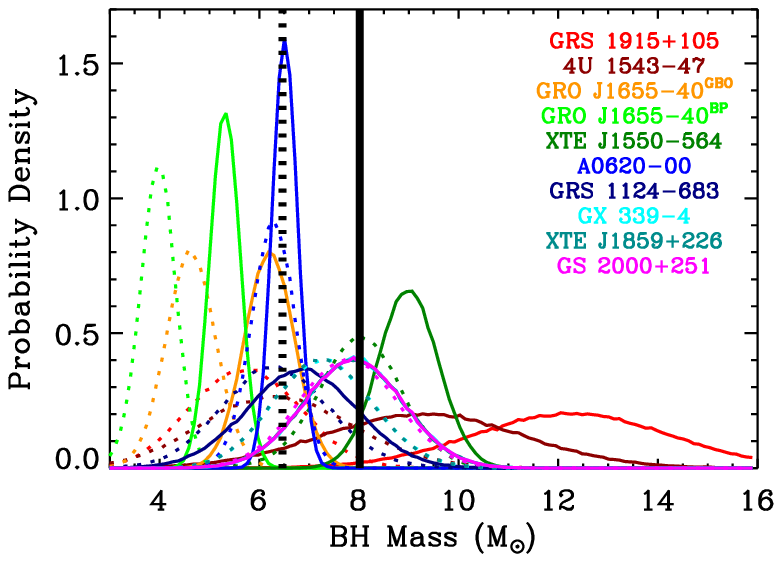}
\caption{\label{BHmass}\textbf{(Left panel):} the evolution of spin parameter $a_*$ with BH mass, as the BH is accumulating mass, assuming again that the birth spin of the BH is zero. For any pair of values of the current BH mass and spin, one can estimate the birth mass of the BH following the appropriate track. The measured BH masses and spins of the 9 LMXBs from Table~\ref{ObsProp} are also shown for comparison. \textbf{(Right panel):} The solid curves show the probability density function for the currently observed mass for the 9 LMXBs systems in Table~\ref{ObsProp}, while the dotted curves correspond to estimated birth BH masses for each system, under the hypothesis that BHs in LMXBs acquire their spin through accretion. The solid vertical line shows the mean currently observed BH mass for the 9 systems in Table~\ref{ObsProp}, while the vertical dotted line is the estimated mean birth BH mass for the same 9 systems. In the Monte-Carlo simulation performed in order to derive the estimated birth BH-mass probability density functions we assumed that the observational errors in the measurement of both the BH mass and BH spin are Gaussian. }
\end{figure*}

If our hypothesis that BHs in Galactic LMXBs acquire their spin through
the accretion of matter during the XRB phase is correct, then this also
implies that the birth mass of BHs that are currently observed to have
high spin can be significantly less than their present-day
mass. Furthermore, the birth mass of a BH in a Galactic LMXB can be
fully determined by combining the present-day measurements of its mass
and spin. The left panel of Figure~\ref{BHmass}, assuming again
that a BH is born with negligible spin, shows the evolution of the spin
as the BH accumulates mass, for different initial BH masses. In this
same figure, the current BH mass and spin for the 9 Galactic BH LMXBs
with measured or estimated BH spins (see Table~\ref{ObsProp}) are
over-plotted. Based on this figure, one can infer the birth mass of the
BH in a LMXB, using the measurements of the current mass and spin of the
BH, by following the appropriate spin evolution track back to spin
zero. For instance, the birth mass of the BH in GRS\,1915+105, which
now has a spin parameter $a_*\sim 1$, has to be about half of its
currently observed mass. However, this figure does not contain any
information on whether a binary evolutionary track that would transfer
the necessary mass from the companion star onto the BH, while also
satisfying all the other currently observed properties of the system,
can exist. This information was extracted from the grid of MT sequences
as described in the previous sections. This procedure allows us to
translate the currently observed BH mass spectrum to the initial mass
function of BHs, which is a crucial step towards understanding BH
formation.

Assuming that the errors in the observationally determined values of
both the present-day masses and spins of the 9 systems in
Table~\ref{ObsProp} are Gaussian and that the BH spin was acquired via
accretion, we performed a simple Monte-Carlo simulation in order to
estimate the birth BH mass for each of these
systems. The right panel of figure~\ref{BHmass} shows the probability density function of
the present-day BH mass (solid lines) for each of the BH LMXBs we
considered, and the estimated, through Monte-Carlo simulations,
probability density function (dotted lines) of their birth BH mass. In
systems like A0620-00, where the measured BH spin is very low, this
process produces just a wider probability density function for the birth
BH mass compared to the present-day BH mass measurement. However, for
LMXBs with high BH spin the probability density functions for the birth
and the present-day BH mass can differ significantly. The most extreme
case is GRS\,1915+105 whose probability density function for the birth
BH mass peaks at $\sim 5\,\rm M_\odot$, while the one for the
present-day BH mass is centered at $\sim 10\,\rm M_\odot$. In the same
figure, the thick solid vertical line shows the mean present-day BH mass
of the nine considered LMXB systems, while the dotted thick vertical
line shows the mean birth mass. The two means differ by $\sim 1.5\,\rm
M_\odot$, while at the same time the standard deviation of the nine
highest likelihood values for the birth BH mass is also slightly smaller
compared to the present day BH mass measurements ($1.6\,\rm M_\odot$
compared to $2.2\,\rm M_\odot$).

\citet{Ozel2010}, using the dynamical, present-day mass measurements of
16 BHs in transient LMXBs inferred that the stellar BH mass
distribution in this subclass of systems is best described by a narrow
distribution at $7.8\pm1.2\,\rm M_{\odot}$. \citet{Farr2011}, using a
bayesian approach, and \citet{Kreidberg2012}, taking into account
potential systematic errors in the BH mass measurement, arrived at
similar conclusions. All three aforementioned studies do not
differentiate between the current and birth BH mass
distribution. However, our analysis strongly indicates that the origin
of BH spin in Galactic LMXBs is the material that the BHs accreted after
their formation, during the XRB phase. Hence, for any BH with
non-negligible spin one can estimate the birth mass of the BH based on
the currently observed mass and spin. Based on these findings, we
conclude that the birth BH mass distribution is shifted towards lower
masses, compared to the currently observed one, and it can be described
by Gaussian distribution at $6.6\pm 1.5\,\rm M_{\odot}$. We should
stress here that the BH mass distribution reported here, and in the
three earlier studies, should not be directly translated to a general BH
mass distribution. It is valid only for BHs found in LMXBs, as the
progenitor stars of these BHs have most likely lost their envelopes due
to binary interactions early in their lives. Hence the mass of the
resulting BH can be significantly lower compared to the BH that a single
isolated star of the same ZAMS mass would produce.

The new derived birth BH mass distribution narrows somewhat, but does
not close, the observed gap between the most massive neutron stars and
the least massive BHs. Our results are consistent with earlier studies
that attempt to explain the existence of a gap in the mass distribution
of compact objects by either adopting the ``rapid'' supernova explosion
mechanism \citep{Fryer2012, Belczynski2012} or by assuming that progenitor
stars in the mass range $16.5\,\rm M_{\odot}\lesssim M \lesssim 25 \,
M_{\odot}$ die in failed supernovae creating BH
\citep{Kochanek2014,Kochanek2014b}, but they do not favor one scenario
over the other. Furthermore, our derived birth-mass distribution for BHs
found in LMXBs separates them even more from BHs found in HMXBs, which
have typical masses $\gtrsim 10\rm \, M_{\odot}$. In the latter class of
XRBs, due to their short lifetimes, BHs do not have a chance to accrete
any significant amount of mass and thus their currently observed mass is
practically the same as their birth mass. The fact that the birth-mass
distribution of BHs in LMXBs seems to have a rapidly declining high-end
mass at $\sim 9\,\rm M_{\odot}$ while the mass spectrum of BHs in HMXBs
extends all the way to $\gtrsim 20\rm \, M_{\odot}$ indicates that the
two classes of systems have very different formation channels.

\section{Summary and Conclusions}

This paper addresses questions related to the origin of BH spin in Galactic LMXBs. Based on the standard formation channel of LMXBs in the Galactic field via evolution of isolated binaries, we argue that at solar-like metallicities, the BH progenitor star will lose during its evolution any significant angular momentum it might initially have had, giving birth to a BH with negligible spin. Therefore, the currently observed BH spin in Galactic LMXBs must be only an effect of mass accretion during the XRB phase. In order to test this hypothesis, we created a grid of $\sim 28,000$ MT sequences, for a BH and its companion star, and used the grid in order to estimate the progenitor properties of currently observed Galactic LMXBs at the onset of RLO. As a benchmark, we compiled a sample of 16 Galactic LMXBs with dynamically confirmed BH accretors. Estimating the progenitor properties of a currently observed LMXB puts constraints on the amount of material that the BH can have accreted, and hence on the spin that the BH can have achieved. Using our grid of MT sequences and the observational sample of 16 Galactic BH LMXBs whose spins have not yet been measured, we were able to test our hypothesis and make robust predictions about the maximum possible spin that the BH in each LMXB can have. The main conclusions of this work can be summarized as follows.

\begin{enumerate}

\item For \emph{all} 16 Galactic LMXBs we were able to find MT sequences that represent possible progenitors of the considered XRBs, and therefore we were able to estimate properties that their progenitors had at the onset of RLO. These ``successful'' MT sequences for each observed system yield possible donor and BH masses and orbital periods at the onset of the MT phase, as well as the maximum amount of material that has been accreted onto the BH since its birth.

\item Based on our estimate of the maximum amount of material that has been accreted since the BH was born, we calculated the maximum spin that the BH can have and compared these values to the BH spin measurements of 5 Galactic LMXBs. We made the same comparison for 4 additional Galactic LMXBs whose spins were estimated via an empirical correlation between the spin parameter and the power of the ballistic jets observed for these systems. In \emph{all} cases we found that the measured or estimated BH spin can be fully accounted by the amount of mass that the BH may have accreted during the MT phase. Thus we conclude that our hypothesis is viable: Galactic LMXBs were born with negligible spin, and that their currently observed spin is an effect of mass accretion that occurred after the BH formed.

\item For the 7 out of the 16 Galactic LMXBs of our sample for which no spin measurement or estimate is available, we made robust predictions about the maximum spin of the BHs they host. These predictions will be tested as new measurements of BH spin become available. Furthermore, for any arbitrary BH LMXB at solar-like metallicity, we are able to set an upper limit on its BH spin based on the other observable properties of the system, such as the orbital period and the effective temperature of the companion star. Specifically, BHs in LMXBs with tight orbits ($P_{\rm orb}\lesssim 0.6$~days), where approximately half of the observed systems are found, can only be mildly spun up to $a_*\lesssim 0.5$ during the MT phase because they could not have accreted more than $1\,\rm M_{\odot}$ of material from their companion star.

\item Our hypothesis that BHs in Galactic LMXBs acquire their spin
  through the accretion of matter during the XRB phase, which is
  supported by the findings of this work, also implies that the birth
  mass of a BH in an LMXB can be significantly less than its present-day
  mass, and moreover, that the birth mass of a BH can be fully
  determined by combining the present-day measurements of its mass and
  spin. Our simple Monte-Carlo simulations show that on average the
  currently measured BH masses in Galactic LMXBs are $\sim1.5\,\rm
  M_{\odot}$ higher than their birth masses. The derived birth BH-mass
  distribution therefore narrows, but does not close, the observed gap
  between the most massive neutron and the least massive BHs. The shift
  downward in the LMXB BH mass distribution also differentiates even
  more the mass spectrum of BHs found in LMXBs to that of BHs found in
  HMXBs, which extends all the way up to $\gtrsim 20\,\rm M_{\odot}$.

\item One additional implication of our hypothesis is that the current
  misalignment of the BH spin compared to the orbital angular momentum
  is expected to be negligible. If the birth BH spin in LMXBs is small,
  as we argue in this study, then the accretion of small a amount of
  material from the companion star will quickly bring the BH spin in
  alignment with the orbit, irrespectively of how large the birth BH
  spin tilt was. Therefore, this work favors the low birth spin
  population synthesis model by \citet{Fragos2010}, which predicts that
  \emph{at least} 95.6\% of BH LMXBs are currently expected to have BH
  spin-orbit misalignments below $20^o$. This finding gives us
  confidence about the robustness of the BH spin measurements via the
  continuum fitting method, which assumes a full alignment between the
  BH spin and the orbital angular momentum.

\item Unlike the case of LMXBs, the spin of BHs found in wind-fed HMXBs must be natal, as the amount of material that these BH can accrete during their short lifetimes is negligible. This result, in conjunction with the significantly different mass spectra for the two categories of BHs, points to distinctly different BH formation channels for LMXBs and HMXBs and different origins for the spins of the BHs they host. In order to understand the origin of BH spin in HMXBs, one needs to study in detail physical processes such as the internal stellar rotation, and the angular momentum exchange between the two stars and the orbit of the progenitor binaries. This will be the subject of future work.

\end{enumerate}

The measurement of the spin of stellar-mass BHs during the last decade opened a new window on our understanding of BH formation and the evolution of BH binaries. This is the first systematic work on the origin of stellar-mass BH spin based on detailed binary calculations. Our findings warrant the continuation of this type of analysis to other classes of XRBs and to other physical processes taking place in the evolution of a binary that affect BH spin.

\acknowledgements We thank R. Narayan for useful discussions and the anonymous referee whose insightful comments improved this paper. TF acknowledges support from the Ambizione Fellowship of the Swiss National Science Foundation (grant PZ00P2\_148123) and the ITC prize fellowship program of the Institute of Theory and Computation at the Harvard-Smithsonian Center for Astrophysics, where part of this work was carried out. JEM acknowledges the support of NASA grant NNX11AD08G. The computations in this paper were run on the Odyssey cluster supported by the FAS Division of Science, Research Computing Group at Harvard University.


\begin{thebibliography}{140}
\expandafter\ifx\csname natexlab\endcsname\relax\def\natexlab#1{#1}\fi

\bibitem[{Axelsson {et~al.}(2011)Axelsson, Church, Davies, Levan, \&
  Ryde}]{Axelsson2011}
Axelsson, M., Church, R.~P., Davies, M.~B., Levan, A.~J., \& Ryde, F. 2011,
  \mnras, 412, 2260

\bibitem[{Bardeen {et~al.}(1972)Bardeen, Press, \& Teukolsky}]{Bardeen1972}
Bardeen, J.~M., Press, W.~H., \& Teukolsky, S.~A. 1972, Astrophysical Journal,
  178, 347

\bibitem[{Barret {et~al.}(1996)Barret, McClintock, \& Grindlay}]{Barret1996}
Barret, D., McClintock, J.~E., \& Grindlay, J.~E. 1996, Astrophysical Journal
  v.473, 473, 963

\bibitem[{Beck {et~al.}(2012)Beck, Montalban, Kallinger, De~Ridder, Aerts,
  Garc{\'\i}a, Hekker, Dupret, Mosser, Eggenberger, Stello, Elsworth, Frandsen,
  Carrier, Hillen, Gruberbauer, Christensen-Dalsgaard, Miglio, Valentini,
  Bedding, Kjeldsen, Girouard, Hall, \& Ibrahim}]{Beck2012}
Beck, P.~G., Montalban, J., Kallinger, T., De~Ridder, J., Aerts, C.,
  Garc{\'\i}a, R.~A., Hekker, S., Dupret, M.-A., Mosser, B., Eggenberger, P.,
  Stello, D., Elsworth, Y., Frandsen, S., Carrier, F., Hillen, M., Gruberbauer,
  M., Christensen-Dalsgaard, J., Miglio, A., Valentini, M., Bedding, T.~R.,
  Kjeldsen, H., Girouard, F.~R., Hall, J.~R., \& Ibrahim, K.~A. 2012, Nature,
  481, 55

\bibitem[{Beer \& Podsiadlowski(2002)}]{BP2002}
Beer, M.~E. \& Podsiadlowski, P. 2002, \mnras, 331, 351

\bibitem[{Belczynski {et~al.}(2010)Belczynski, Bulik, Fryer, Ruiter, Valsecchi,
  Vink, \& Hurley}]{Belczynski2010}
Belczynski, K., Bulik, T., Fryer, C.~L., Ruiter, A., Valsecchi, F., Vink,
  J.~S., \& Hurley, J.~R. 2010, The Astrophysical Journal, 714, 1217

\bibitem[{Belczynski {et~al.}(2012)Belczynski, Wiktorowicz, Fryer, Holz, \&
  Kalogera}]{Belczynski2012}
Belczynski, K., Wiktorowicz, G., Fryer, C.~L., Holz, D.~E., \& Kalogera, V.
  2012, The Astrophysical Journal, 757, 91

\bibitem[{Belloni {et~al.}(2012)Belloni, Sanna, \& M{\'e}ndez}]{Belloni2012}
Belloni, T.~M., Sanna, A., \& M{\'e}ndez, M. 2012, \mnras, 426, 1701

\bibitem[{Bhattacharya \& van~den Heuvel(1991)}]{BvdH1991}
Bhattacharya, D. \& van~den Heuvel, E. P.~J. 1991, Physics Reports, 203, 1

\bibitem[{Bolton(1972)}]{Bolton1972}
Bolton, C.~T. 1972, Nature Physical Science, 240, 124

\bibitem[{Calvelo {et~al.}(2009)Calvelo, Vrtilek, Steeghs, Torres, Neilsen,
  Filippenko, \& Gonz{\'a}lez~Hern{\'a}ndez}]{Calvelo2009}
Calvelo, D.~E., Vrtilek, S.~D., Steeghs, D., Torres, M. A.~P., Neilsen, J.,
  Filippenko, A.~V., \& Gonz{\'a}lez~Hern{\'a}ndez, J.~I. 2009, \mnras, 399,
  539

\bibitem[{Cantrell {et~al.}(2010)Cantrell, Bailyn, Orosz, McClintock,
  Remillard, Froning, Neilsen, Gelino, \& Gou}]{Cantrell2010}
Cantrell, A.~G., Bailyn, C.~D., Orosz, J.~A., McClintock, J.~E., Remillard,
  R.~A., Froning, C.~S., Neilsen, J., Gelino, D.~M., \& Gou, L. 2010, The
  Astrophysical Journal, 710, 1127

\bibitem[{Casares \& Charles(1994)}]{CC1994}
Casares, J. \& Charles, P.~A. 1994, \mnras, 271, L5

\bibitem[{Casares {et~al.}(1995)Casares, Charles, \& Marsh}]{Casares1995}
Casares, J., Charles, P.~A., \& Marsh, T.~R. 1995, \mnras, 277, L45

\bibitem[{Casares {et~al.}(1997)Casares, Mart{\'\i}n, Charles, Molaro, \&
  Rebolo}]{Casares1997}
Casares, J., Mart{\'\i}n, E.~L., Charles, P.~A., Molaro, P., \& Rebolo, R.
  1997, New Astronomy, 1, 299

\bibitem[{Casares {et~al.}(2009)Casares, Orosz, Zurita, Shahbaz,
  Corral-Santana, McClintock, Garcia, Mart{\'\i}nez-Pais, Charles, Fender, \&
  Remillard}]{Casares2009}
Casares, J., Orosz, J.~A., Zurita, C., Shahbaz, T., Corral-Santana, J.~M.,
  McClintock, J.~E., Garcia, M.~R., Mart{\'\i}nez-Pais, I.~G., Charles, P.~A.,
  Fender, R.~P., \& Remillard, R.~A. 2009, \apjs, 181, 238

\bibitem[{Casares {et~al.}(2004)Casares, Zurita, Shahbaz, Charles, \&
  Fender}]{Casares2004}
Casares, J., Zurita, C., Shahbaz, T., Charles, P.~A., \& Fender, R.~P. 2004,
  The Astrophysical Journal, 613, L133

\bibitem[{Chen(2011)}]{Chen2011}
Chen, T. 2011, Jets at all Scales, 275, 327

\bibitem[{Cowley {et~al.}(1987)Cowley, Crampton, \& Hutchings}]{Cowley1987}
Cowley, A.~P., Crampton, D., \& Hutchings, J.~B. 1987, Astronomical Journal
  (ISSN 0004-6256), 93, 195

\bibitem[{Cox(2000)}]{Cox2000}
Cox, A.~N. 2000, {Allen's astrophysical quantities}, 4th edn. (New York:
  Allen's astrophysical quantities)

\bibitem[{Detmers {et~al.}(2008)Detmers, Langer, Podsiadlowski, \&
  Izzard}]{Detmers2008}
Detmers, R.~G., Langer, N., Podsiadlowski, P., \& Izzard, R.~G. 2008, \aap,
  484, 831

\bibitem[{Dexter \& Blaes(2014)}]{DB2014}
Dexter, J. \& Blaes, O. 2014, \mnras, 438, 3352

\bibitem[{Dubus {et~al.}(1999)Dubus, Lasota, Hameury, \& Charles}]{Dubus1999}
Dubus, G., Lasota, J.-P., Hameury, J.-M., \& Charles, P. 1999, \mnras, 303, 139

\bibitem[{Fabian {et~al.}(1989)Fabian, Rees, Stella, \& White}]{Fabian1989}
Fabian, A.~C., Rees, M.~J., Stella, L., \& White, N. 1989, Monthly Notices of
  the Royal Astronomical Society (ISSN 0035-8711), 238, 729

\bibitem[{Farr {et~al.}(2011{\natexlab{a}})Farr, Kremer, Lyutikov, \&
  Kalogera}]{Farr2011b}
Farr, W.~M., Kremer, K., Lyutikov, M., \& Kalogera, V. 2011{\natexlab{a}}, The
  Astrophysical Journal, 742, 81

\bibitem[{Farr {et~al.}(2011{\natexlab{b}})Farr, Sravan, Cantrell, Kreidberg,
  Bailyn, Mandel, \& Kalogera}]{Farr2011}
Farr, W.~M., Sravan, N., Cantrell, A., Kreidberg, L., Bailyn, C.~D., Mandel,
  I., \& Kalogera, V. 2011{\natexlab{b}}, The Astrophysical Journal, 741, 103

\bibitem[{Filippenko \& Chornock(2001)}]{FC2001}
Filippenko, A.~V. \& Chornock, R. 2001, IAU Circ., 7644, 2

\bibitem[{Filippenko {et~al.}(1999)Filippenko, Leonard, Matheson, Li, Moran, \&
  Riess}]{Filippenko1999}
Filippenko, A.~V., Leonard, D.~C., Matheson, T., Li, W., Moran, E.~C., \&
  Riess, A.~G. 1999, The Publications of the Astronomical Society of the
  Pacific, 111, 969

\bibitem[{Filippenko {et~al.}(1995)Filippenko, Matheson, \&
  Barth}]{Filippenko1995}
Filippenko, A.~V., Matheson, T., \& Barth, A.~J. 1995, Astrophysical Journal
  Letters v.455, 455, L139

\bibitem[{Fragos {et~al.}(2010)Fragos, Tremmel, Rantsiou, \&
  Belczynski}]{Fragos2010}
Fragos, T., Tremmel, M., Rantsiou, E., \& Belczynski, K. 2010, \apjl, 719, L79

\bibitem[{Fragos {et~al.}(2009)Fragos, Willems, Kalogera, Ivanova, Rockefeller,
  Fryer, \& Young}]{Fragos2009b}
Fragos, T., Willems, B., Kalogera, V., Ivanova, N., Rockefeller, G., Fryer,
  C.~L., \& Young, P.~A. 2009, The Astrophysical Journal, 697, 1057

\bibitem[{Fryer {et~al.}(2012)Fryer, Belczynski, Wiktorowicz, Dominik,
  Kalogera, \& Holz}]{Fryer2012}
Fryer, C.~L., Belczynski, K., Wiktorowicz, G., Dominik, M., Kalogera, V., \&
  Holz, D.~E. 2012, The Astrophysical Journal, 749, 91

\bibitem[{Gelino(2004)}]{Gelino2004}
Gelino, D.~M. 2004, Compact Binaries in the Galaxy and Beyond, 20, 214

\bibitem[{Gelino {et~al.}(2001)Gelino, Harrison, \& McNamara}]{Gelino2001}
Gelino, D.~M., Harrison, T.~E., \& McNamara, B.~J. 2001, The Astronomical
  Journal, 122, 971

\bibitem[{Georgy {et~al.}(2009)Georgy, Meynet, Walder, Folini, \&
  Maeder}]{Georgy2009}
Georgy, C., Meynet, G., Walder, R., Folini, D., \& Maeder, A. 2009, \aap, 502,
  611

\bibitem[{Gonz{\'a}lez~Hern{\'a}ndez {et~al.}(2012)Gonz{\'a}lez~Hern{\'a}ndez,
  Rebolo, \& Casares}]{Gonzalez2012}
Gonz{\'a}lez~Hern{\'a}ndez, J.~I., Rebolo, R., \& Casares, J. 2012, \apjl, 744,
  L25

\bibitem[{Gou {et~al.}(2009)Gou, McClintock, Liu, Narayan, Steiner, Remillard,
  Orosz, Davis, Ebisawa, \& Schlegel}]{Gou2009}
Gou, L., McClintock, J.~E., Liu, J., Narayan, R., Steiner, J.~F., Remillard,
  R.~A., Orosz, J.~A., Davis, S.~W., Ebisawa, K., \& Schlegel, E.~M. 2009, The
  Astrophysical Journal, 701, 1076

\bibitem[{Gou {et~al.}(2011)Gou, McClintock, Reid, Orosz, Steiner, Narayan,
  Xiang, Remillard, Arnaud, \& Davis}]{Gou2011}
Gou, L., McClintock, J.~E., Reid, M.~J., Orosz, J.~A., Steiner, J.~F., Narayan,
  R., Xiang, J., Remillard, R.~A., Arnaud, K.~A., \& Davis, S.~W. 2011, The
  Astrophysical Journal, 742, 85

\bibitem[{Gou {et~al.}(2014)Gou, McClintock, Remillard, Steiner, Reid, Orosz,
  Narayan, Hanke, \& Garc{\'\i}a}]{Gou2014}
Gou, L., McClintock, J.~E., Remillard, R.~A., Steiner, J.~F., Reid, M.~J.,
  Orosz, J.~A., Narayan, R., Hanke, M., \& Garc{\'\i}a, J. 2014, The
  Astrophysical Journal, 790, 29

\bibitem[{Gou {et~al.}(2010)Gou, McClintock, Steiner, Narayan, Cantrell,
  Bailyn, \& Orosz}]{Gou2010}
Gou, L., McClintock, J.~E., Steiner, J.~F., Narayan, R., Cantrell, A.~G.,
  Bailyn, C.~D., \& Orosz, J.~A. 2010, \apjl, 718, L122

\bibitem[{Gray(2008)}]{Gray2008}
Gray, D.~F. 2008, The Observation and Analysis of Stellar Photospheres

\bibitem[{Greene {et~al.}(2001)Greene, Bailyn, \& Orosz}]{GBO2001}
Greene, J., Bailyn, C.~D., \& Orosz, J.~A. 2001, The Astrophysical Journal,
  554, 1290

\bibitem[{Greiner {et~al.}(2001{\natexlab{a}})Greiner, Cuby, \&
  McCaughrean}]{Greiner2001b}
Greiner, J., Cuby, J.~G., \& McCaughrean, M.~J. 2001{\natexlab{a}}, Nature,
  414, 522

\bibitem[{Greiner {et~al.}(2001{\natexlab{b}})Greiner, Cuby, McCaughrean,
  Castro-Tirado, \& Mennickent}]{Greiner2001}
Greiner, J., Cuby, J.~G., McCaughrean, M.~J., Castro-Tirado, A.~J., \&
  Mennickent, R.~E. 2001{\natexlab{b}}, \aap, 373, L37

\bibitem[{Harlaftis {et~al.}(1999)Harlaftis, Collier, Horne, \&
  Filippenko}]{Harlaftis1999}
Harlaftis, E., Collier, S., Horne, K., \& Filippenko, A.~V. 1999, \aap, 341,
  491

\bibitem[{Harlaftis \& Greiner(2004)}]{HG2004}
Harlaftis, E.~T. \& Greiner, J. 2004, \aap, 414, L13

\bibitem[{Harlaftis {et~al.}(1996)Harlaftis, Horne, \&
  Filippenko}]{Harlaftis1996}
Harlaftis, E.~T., Horne, K., \& Filippenko, A.~V. 1996, Publications of the
  Astronomical Society of the Pacific, 108, 762

\bibitem[{Harlaftis {et~al.}(1997)Harlaftis, Steeghs, Horne, \&
  Filippenko}]{Harlaftis1997}
Harlaftis, E.~T., Steeghs, D., Horne, K., \& Filippenko, A.~V. 1997,
  Astronomical Journal v.114, 114, 1170

\bibitem[{Heger {et~al.}(2005)Heger, Woosley, \& Spruit}]{HWS2005}
Heger, A., Woosley, S.~E., \& Spruit, H.~C. 2005, The Astrophysical Journal,
  626, 350

\bibitem[{Hynes {et~al.}(2009)Hynes, Bradley, Rupen, Gallo, Fender, Casares, \&
  Zurita}]{Hynes2009}
Hynes, R.~I., Bradley, C.~K., Rupen, M., Gallo, E., Fender, R.~P., Casares, J.,
  \& Zurita, C. 2009, \mnras, 399, 2239

\bibitem[{Hynes {et~al.}(2003)Hynes, Steeghs, Casares, Charles, \&
  O'Brien}]{Hynes2003}
Hynes, R.~I., Steeghs, D., Casares, J., Charles, P.~A., \& O'Brien, K. 2003,
  The Astrophysical Journal, 583, L95

\bibitem[{Ivanova {et~al.}(2012)Ivanova, Justham, Chen, De~Marco, Fryer,
  Gaburov, Ge, Glebbeek, Han, Li, Lu, Marsh, Podsiadlowski, Potter, Soker,
  Taam, Tauris, van~den Heuvel, \& Webbink}]{Ivanova2012b}
Ivanova, N., Justham, S., Chen, X., De~Marco, O., Fryer, C.~L., Gaburov, E.,
  Ge, H., Glebbeek, E., Han, Z., Li, X.~D., Lu, G., Marsh, T., Podsiadlowski,
  P., Potter, A., Soker, N., Taam, R., Tauris, T.~M., van~den Heuvel, E. P.~J.,
  \& Webbink, R.~F. 2012, arXiv, 4302

\bibitem[{Ivanova {et~al.}(2002)Ivanova, Podsiadlowski, \& Spruit}]{IPS2002}
Ivanova, N., Podsiadlowski, P., \& Spruit, H. 2002, \mnras, 334, 819

\bibitem[{Izzard {et~al.}(2004)Izzard, Ramirez-Ruiz, \& Tout}]{Izzard2004}
Izzard, R.~G., Ramirez-Ruiz, E., \& Tout, C.~A. 2004, \mnras, 348, 1215

\bibitem[{Johannsen {et~al.}(2009)Johannsen, Psaltis, \&
  McClintock}]{Johannsen2009}
Johannsen, T., Psaltis, D., \& McClintock, J.~E. 2009, The Astrophysical
  Journal, 691, 997

\bibitem[{Kawaler(1988)}]{Kawaler1988}
Kawaler, S.~D. 1988, Astrophysical Journal, 333, 236

\bibitem[{Khargharia {et~al.}(2010)Khargharia, Froning, \&
  Robinson}]{Khargharia2010}
Khargharia, J., Froning, C.~S., \& Robinson, E.~L. 2010, The Astrophysical
  Journal, 716, 1105

\bibitem[{Khargharia {et~al.}(2013)Khargharia, Froning, Robinson, \&
  Gelino}]{Khargharia2013}
Khargharia, J., Froning, C.~S., Robinson, E.~L., \& Gelino, D.~M. 2013, The
  Astronomical Journal, 145, 21

\bibitem[{King {et~al.}(1996)King, Kolb, \& Burderi}]{KKB1996}
King, A.~R., Kolb, U., \& Burderi, L. 1996, Astrophysical Journal Letters
  v.464, 464, L127

\bibitem[{Kochanek(2014{\natexlab{a}})}]{Kochanek2014b}
Kochanek, C.~S. 2014{\natexlab{a}}, arXiv, 5622

\bibitem[{Kochanek(2014{\natexlab{b}})}]{Kochanek2014}
---. 2014{\natexlab{b}}, The Astrophysical Journal, 785, 28

\bibitem[{Kreidberg {et~al.}(2012)Kreidberg, Bailyn, Farr, \&
  Kalogera}]{Kreidberg2012}
Kreidberg, L., Bailyn, C.~D., Farr, W.~M., \& Kalogera, V. 2012, The
  Astrophysical Journal, 757, 36

\bibitem[{Kulkarni {et~al.}(2011)Kulkarni, Penna, Shcherbakov, Steiner,
  Narayan, S{\"a}~Dowski, Zhu, McClintock, Davis, \& McKinney}]{Kulkarni2011}
Kulkarni, A.~K., Penna, R.~F., Shcherbakov, R.~V., Steiner, J.~F., Narayan, R.,
  S{\"a}~Dowski, A., Zhu, Y., McClintock, J.~E., Davis, S.~W., \& McKinney,
  J.~C. 2011, \mnras, 414, 1183

\bibitem[{Langer(2012)}]{Langer2012}
Langer, N. 2012, \araa, 50, 107

\bibitem[{Lee {et~al.}(2002)Lee, Brown, \& Wijers}]{LBW2002}
Lee, C.-H., Brown, G.~E., \& Wijers, R. A. M.~J. 2002, The Astrophysical
  Journal, 575, 996

\bibitem[{Liu {et~al.}(2008)Liu, McClintock, Narayan, Davis, \&
  Orosz}]{Liu2008}
Liu, J., McClintock, J.~E., Narayan, R., Davis, S.~W., \& Orosz, J.~A. 2008,
  The Astrophysical Journal, 679, L37

\bibitem[{Liu {et~al.}(2010)Liu, McClintock, Narayan, Davis, \&
  Orosz}]{Liu2010}
---. 2010, \apjl, 719, L109

\bibitem[{MacDonald {et~al.}(2014)MacDonald, Bailyn, Buxton, Cantrell,
  Chatterjee, Kennedy-Shaffer, Orosz, Markwardt, \& Swank}]{MacDonald2014}
MacDonald, R. K.~D., Bailyn, C.~D., Buxton, M., Cantrell, A.~G., Chatterjee,
  R., Kennedy-Shaffer, R., Orosz, J.~A., Markwardt, C.~B., \& Swank, J.~H.
  2014, The Astrophysical Journal, 784, 2

\bibitem[{Macias {et~al.}(2011)Macias, Orosz, Bailyn, Buxton, Schechter,
  Remillard, McClintock, \& Steiner}]{Macias2011}
Macias, P., Orosz, J.~A., Bailyn, C.~D., Buxton, M.~M., Schechter, P.~L.,
  Remillard, R.~A., McClintock, J.~E., \& Steiner, J.~F. 2011, American
  Astronomical Society, 217, 14304

\bibitem[{Maeder \& Meynet(2012)}]{MM2012}
Maeder, A. \& Meynet, G. 2012, Reviews of Modern Physics, 84, 25

\bibitem[{McClintock {et~al.}(2013)McClintock, Narayan, \&
  Steiner}]{McClintock2013}
McClintock, J.~E., Narayan, R., \& Steiner, J.~F. 2013, Space Science Reviews,
  73

\bibitem[{McClintock \& Remillard(2006)}]{McCR2006}
McClintock, J.~E. \& Remillard, R.~A. 2006, In: Compact stellar X-ray sources.
  Edited by Walter Lewin {\&} Michiel van der Klis. Cambridge Astrophysics
  Series, 157

\bibitem[{McClintock {et~al.}(2006)McClintock, Shafee, Narayan, Remillard,
  Davis, \& Li}]{McClintock2006}
McClintock, J.~E., Shafee, R., Narayan, R., Remillard, R.~A., Davis, S.~W., \&
  Li, L.-X. 2006, The Astrophysical Journal, 652, 518

\bibitem[{Menou {et~al.}(2002)Menou, Perna, \& Hernquist}]{MPH2002}
Menou, K., Perna, R., \& Hernquist, L. 2002, The Astrophysical Journal, 564,
  L81

\bibitem[{Meynet {et~al.}(2008)Meynet, Walborn, Hunter, Martayan, van Marle,
  Marchenko, Vink, Limongi, Levesque, \& Modjaz}]{Meynet2008}
Meynet, G., Walborn, N.~R., Hunter, I., Martayan, C., van Marle, A.~J.,
  Marchenko, S., Vink, J.~S., Limongi, M., Levesque, E.~M., \& Modjaz, M. 2008,
  Massive Stars as Cosmic Engines, 250, 571

\bibitem[{Moreno~M{\'e}ndez(2011)}]{Mendez2011b}
Moreno~M{\'e}ndez, E. 2011, \mnras, 413, 183

\bibitem[{Moreno~M{\'e}ndez {et~al.}(2008)Moreno~M{\'e}ndez, Brown, Lee, \&
  Park}]{Mendez2008}
Moreno~M{\'e}ndez, E., Brown, G.~E., Lee, C.-H., \& Park, I.~H. 2008, The
  Astrophysical Journal, 689, L9

\bibitem[{Moreno~M{\'e}ndez {et~al.}(2011)Moreno~M{\'e}ndez, Brown, Lee, \&
  Walter}]{Mendez2011a}
Moreno~M{\'e}ndez, E., Brown, G.~E., Lee, C.-H., \& Walter, F.~M. 2011, The
  Astrophysical Journal, 727, 29

\bibitem[{Morningstar {et~al.}(2014)Morningstar, Miller, Reis, \&
  Ebisawa}]{Morningstar2014}
Morningstar, W.~R., Miller, J.~M., Reis, R.~C., \& Ebisawa, K. 2014, \apjl,
  784, L18

\bibitem[{Motta {et~al.}(2014{\natexlab{a}})Motta, Belloni, Stella,
  Mu{\~n}oz-Darias, \& Fender}]{Motta2014}
Motta, S.~E., Belloni, T.~M., Stella, L., Mu{\~n}oz-Darias, T., \& Fender, R.
  2014{\natexlab{a}}, \mnras, 437, 2554

\bibitem[{Motta {et~al.}(2014{\natexlab{b}})Motta, Mu{\~n}oz-Darias, Sanna,
  Fender, Belloni, \& Stella}]{Motta2014b}
Motta, S.~E., Mu{\~n}oz-Darias, T., Sanna, A., Fender, R., Belloni, T., \&
  Stella, L. 2014{\natexlab{b}}, Monthly Notices of the Royal Astronomical
  Society: Letters, 439, L65

\bibitem[{Naoz \& Fabrycky(2014)}]{Naoz2014}
Naoz, S. \& Fabrycky, D.~C. 2014, arXiv, 5223

\bibitem[{Naoz {et~al.}(2011)Naoz, Farr, Lithwick, Rasio, \&
  Teyssandier}]{Naoz2011}
Naoz, S., Farr, W.~M., Lithwick, Y., Rasio, F.~A., \& Teyssandier, J. 2011,
  Nature, 473, 187

\bibitem[{Naoz {et~al.}(2013)Naoz, Farr, Lithwick, Rasio, \&
  Teyssandier}]{Naoz2013}
---. 2013, \mnras, 431, 2155

\bibitem[{Narayan \& McClintock(2012)}]{NMcC2012}
Narayan, R. \& McClintock, J.~E. 2012, Monthly Notices of the Royal
  Astronomical Society: Letters, 419, L69

\bibitem[{Neilsen {et~al.}(2008)Neilsen, Steeghs, \& Vrtilek}]{NSV2008}
Neilsen, J., Steeghs, D., \& Vrtilek, S.~D. 2008, \mnras, 384, 849

\bibitem[{Noble {et~al.}(2011)Noble, Krolik, Schnittman, \& Hawley}]{Noble2011}
Noble, S.~C., Krolik, J.~H., Schnittman, J.~D., \& Hawley, J.~F. 2011, The
  Astrophysical Journal, 743, 115

\bibitem[{Novikov \& Thorne(1973)}]{NT1973}
Novikov, I.~D. \& Thorne, K.~S. 1973, Black holes (Les astres occlus), 343

\bibitem[{Orosz(2003)}]{Orosz2003}
Orosz, J.~A. 2003, A Massive Star Odyssey: From Main Sequence to Supernova,
  212, 365

\bibitem[{Orosz {et~al.}(1996)Orosz, Bailyn, McClintock, \&
  Remillard}]{Orosz1996}
Orosz, J.~A., Bailyn, C.~D., McClintock, J.~E., \& Remillard, R.~A. 1996,
  Astrophysical Journal v.468, 468, 380

\bibitem[{Orosz {et~al.}(1998)Orosz, Jain, Bailyn, McClintock, \&
  Remillard}]{Orosz1998}
Orosz, J.~A., Jain, R.~K., Bailyn, C.~D., McClintock, J.~E., \& Remillard,
  R.~A. 1998, Astrophysical Journal v.499, 499, 375

\bibitem[{Orosz {et~al.}(2001)Orosz, Kuulkers, van~der Klis, McClintock,
  Garcia, Callanan, Bailyn, Jain, \& Remillard}]{Orosz2001}
Orosz, J.~A., Kuulkers, E., van~der Klis, M., McClintock, J.~E., Garcia, M.~R.,
  Callanan, P.~J., Bailyn, C.~D., Jain, R.~K., \& Remillard, R.~A. 2001, The
  Astrophysical Journal, 555, 489

\bibitem[{Orosz {et~al.}(2011{\natexlab{a}})Orosz, McClintock, Aufdenberg,
  Remillard, Reid, Narayan, \& Gou}]{2011ApJ...742...84O}
Orosz, J.~A., McClintock, J.~E., Aufdenberg, J.~P., Remillard, R.~A., Reid,
  M.~J., Narayan, R., \& Gou, L. 2011{\natexlab{a}}, The Astrophysical Journal,
  742, 84

\bibitem[{Orosz {et~al.}(2014)Orosz, Steiner, McClintock, Buxton, Bailyn,
  Steeghs, Guberman, \& Torres}]{Orosz2014}
Orosz, J.~A., Steiner, J.~F., McClintock, J.~E., Buxton, M.~M., Bailyn, C.~D.,
  Steeghs, D., Guberman, A., \& Torres, M. A.~P. 2014, arXiv, 85

\bibitem[{Orosz {et~al.}(2011{\natexlab{b}})Orosz, Steiner, McClintock, Torres,
  Remillard, Bailyn, \& Miller}]{Orosz2011}
Orosz, J.~A., Steiner, J.~F., McClintock, J.~E., Torres, M. A.~P., Remillard,
  R.~A., Bailyn, C.~D., \& Miller, J.~M. 2011{\natexlab{b}}, The Astrophysical
  Journal, 730, 75

\bibitem[{{\"O}zel {et~al.}(2010){\"O}zel, Psaltis, Narayan, \&
  McClintock}]{Ozel2010}
{\"O}zel, F., Psaltis, D., Narayan, R., \& McClintock, J.~E. 2010, The
  Astrophysical Journal, 725, 1918

\bibitem[{Paczy{\'{n}}ski(1971)}]{Paczynski1971}
Paczy{\'{n}}ski, B. 1971, Acta Astronomica, 21, 1

\bibitem[{Paxton {et~al.}(2011)Paxton, Bildsten, Dotter, Herwig, Lesaffre, \&
  Timmes}]{Paxton2011}
Paxton, B., Bildsten, L., Dotter, A., Herwig, F., Lesaffre, P., \& Timmes, F.
  2011, \apjs, 192, 3

\bibitem[{Paxton {et~al.}(2013)Paxton, Cantiello, Arras, Bildsten, Brown,
  Dotter, Mankovich, Montgomery, Stello, Timmes, \& Townsend}]{Paxton2013}
Paxton, B., Cantiello, M., Arras, P., Bildsten, L., Brown, E.~F., Dotter, A.,
  Mankovich, C., Montgomery, M.~H., Stello, D., Timmes, F.~X., \& Townsend, R.
  2013, arXiv, 319

\bibitem[{Penna {et~al.}(2010)Penna, McKinney, Narayan, Tchekhovskoy, Shafee,
  \& McClintock}]{Penna2010}
Penna, R.~F., McKinney, J.~C., Narayan, R., Tchekhovskoy, A., Shafee, R., \&
  McClintock, J.~E. 2010, \mnras, 408, 752

\bibitem[{Podsiadlowski {et~al.}(2004)Podsiadlowski, Mazzali, Nomoto, Lazzati,
  \& Cappellaro}]{Podsiadlowski2004}
Podsiadlowski, P., Mazzali, P.~A., Nomoto, K., Lazzati, D., \& Cappellaro, E.
  2004, The Astrophysical Journal, 607, L17

\bibitem[{Podsiadlowski {et~al.}(2003)Podsiadlowski, Rappaport, \&
  Han}]{PRH2003}
Podsiadlowski, P., Rappaport, S., \& Han, Z. 2003, Monthly Notice of the Royal
  Astronomical Society, 341, 385

\bibitem[{Rappaport {et~al.}(1995)Rappaport, Podsiadlowski, {Joss, P. C.},
  Di~Stefano, \& Han}]{Rappaport1995}
Rappaport, S., Podsiadlowski, P., {Joss, P. C.}, Di~Stefano, R., \& Han, Z.
  1995, \mnras, 273, 731

\bibitem[{Reid {et~al.}(2014)Reid, McClintock, Steiner, Steeghs, Remillard,
  Dhawan, \& Narayan}]{Reid2014}
Reid, M.~J., McClintock, J.~E., Steiner, J.~F., Steeghs, D., Remillard, R.~A.,
  Dhawan, V., \& Narayan, R. 2014, arXiv, 2453

\bibitem[{Remillard \& McClintock(2006)}]{RMcC2006}
Remillard, R.~A. \& McClintock, J.~E. 2006, \araa, 44, 49

\bibitem[{Remillard {et~al.}(1996)Remillard, Orosz, McClintock, \&
  Bailyn}]{Remillard1996}
Remillard, R.~A., Orosz, J.~A., McClintock, J.~E., \& Bailyn, C.~D. 1996,
  Astrophysical Journal v.459, 459, 226

\bibitem[{Repetto {et~al.}(2012)Repetto, Davies, \& Sigurdsson}]{Repetto2012}
Repetto, S., Davies, M.~B., \& Sigurdsson, S. 2012, \mnras, 425, 2799

\bibitem[{Reynolds(2013)}]{Reynolds2013}
Reynolds, C.~S. 2013, Space Science Reviews, 183, 81

\bibitem[{Reynolds \& Miller(2009)}]{RM2009}
Reynolds, C.~S. \& Miller, M.~C. 2009, The Astrophysical Journal, 692, 869

\bibitem[{Russell {et~al.}(2013)Russell, Gallo, \& Fender}]{RGF2013}
Russell, D.~M., Gallo, E., \& Fender, R.~P. 2013, \mnras, 431, 405

\bibitem[{Sadakane {et~al.}(2006)Sadakane, Arai, Aoki, Arimoto, Takada-Hidai,
  Ohnishi, Tajitsu, Beers, Iwamoto, Tominaga, Umeda, Maeda, \&
  Nomoto}]{Sadakane2006}
Sadakane, K., Arai, A., Aoki, W., Arimoto, N., Takada-Hidai, M., Ohnishi, T.,
  Tajitsu, A., Beers, T.~C., Iwamoto, N., Tominaga, N., Umeda, H., Maeda, K.,
  \& Nomoto, K. 2006, Publications of the Astronomical Society of Japan, 58,
  595

\bibitem[{Shafee {et~al.}(2006)Shafee, McClintock, Narayan, Davis, Li, \&
  Remillard}]{Shafee2006}
Shafee, R., McClintock, J.~E., Narayan, R., Davis, S.~W., Li, L.-X., \&
  Remillard, R.~A. 2006, The Astrophysical Journal, 636, L113

\bibitem[{Shafee {et~al.}(2008)Shafee, McKinney, Narayan, Tchekhovskoy, Gammie,
  \& McClintock}]{Shafee2008}
Shafee, R., McKinney, J.~C., Narayan, R., Tchekhovskoy, A., Gammie, C.~F., \&
  McClintock, J.~E. 2008, The Astrophysical Journal, 687, L25

\bibitem[{Shahbaz {et~al.}(1999)Shahbaz, van~der Hooft, Casares, Charles, \&
  van Paradijs}]{Shahbaz1999}
Shahbaz, T., van~der Hooft, F., Casares, J., Charles, P.~A., \& van Paradijs,
  J. 1999, \mnras, 306, 89

\bibitem[{Shakura \& Sunyaev(1973)}]{SS1973}
Shakura, N.~I. \& Sunyaev, R.~A. 1973, \aap, 24, 337

\bibitem[{Steeghs {et~al.}(2013)Steeghs, McClintock, Parsons, Reid, Littlefair,
  \& Dhillon}]{Steeghs2013}
Steeghs, D., McClintock, J.~E., Parsons, S.~G., Reid, M.~J., Littlefair, S., \&
  Dhillon, V.~S. 2013, The Astrophysical Journal, 768, 185

\bibitem[{Steiner {et~al.}(2013)Steiner, McClintock, \& Narayan}]{Steiner2013}
Steiner, J.~F., McClintock, J.~E., \& Narayan, R. 2013, The Astrophysical
  Journal, 762, 104

\bibitem[{Steiner {et~al.}(2014{\natexlab{a}})Steiner, McClintock, Orosz,
  Remillard, Bailyn, Kolehmainen, \& Straub}]{Steiner:2014dx}
Steiner, J.~F., McClintock, J.~E., Orosz, J.~A., Remillard, R.~A., Bailyn,
  C.~D., Kolehmainen, M., \& Straub, O. 2014{\natexlab{a}}, \apj, 793, L29

\bibitem[{Steiner {et~al.}(2014{\natexlab{b}})Steiner, McClintock, Orosz,
  Remillard, Bailyn, Kolehmainen, \& Straub}]{Steiner2014}
---. 2014{\natexlab{b}}, arXiv, 148

\bibitem[{Steiner {et~al.}(2012)Steiner, McClintock, \& Reid}]{Steiner2012}
Steiner, J.~F., McClintock, J.~E., \& Reid, M.~J. 2012, \apjl, 745, L7

\bibitem[{Steiner {et~al.}(2011)Steiner, Reis, McClintock, Narayan, Remillard,
  Orosz, Gou, Fabian, \& Torres}]{Steiner2011}
Steiner, J.~F., Reis, R.~C., McClintock, J.~E., Narayan, R., Remillard, R.~A.,
  Orosz, J.~A., Gou, L., Fabian, A.~C., \& Torres, M. A.~P. 2011, \mnras, 416,
  941

\bibitem[{Suijs {et~al.}(2008)Suijs, Langer, Poelarends, Yoon, Heger, \&
  Herwig}]{Suijs2008}
Suijs, M. P.~L., Langer, N., Poelarends, A.-J., Yoon, S.-C., Heger, A., \&
  Herwig, F. 2008, \aap, 481, L87

\bibitem[{Taam \& Ricker(2010)}]{TR2010}
Taam, R.~E. \& Ricker, P.~M. 2010, New Astronomy Reviews, 54, 65

\bibitem[{Tauris \& van~den Heuvel(2006)}]{TvdH2006}
Tauris, T.~M. \& van~den Heuvel, E. P.~J. 2006, In: Compact stellar X-ray
  sources. Edited by Walter Lewin {\&} Michiel van der Klis. Cambridge
  Astrophysics Series, 623

\bibitem[{Thorne(1974)}]{Thorne1974}
Thorne, K.~S. 1974, Astrophysical Journal, 191, 507

\bibitem[{T{\"o}r{\"o}k {et~al.}(2011)T{\"o}r{\"o}k, Kotrlov{\'a},
  {\v{S}}r{\'a}mkov{\'a}, \& Stuchl{\'\i}k}]{Torok2011}
T{\"o}r{\"o}k, G., Kotrlov{\'a}, A., {\v{S}}r{\'a}mkov{\'a}, E., \&
  Stuchl{\'\i}k, Z. 2011, \aap, 531, 59

\bibitem[{Valsecchi {et~al.}(2010)Valsecchi, Glebbeek, Farr, Fragos, Willems,
  Orosz, Liu, \& Kalogera}]{Valsecchi2010}
Valsecchi, F., Glebbeek, E., Farr, W.~M., Fragos, T., Willems, B., Orosz,
  J.~A., Liu, J., \& Kalogera, V. 2010, Nature, 468, 77

\bibitem[{van~den Heuvel \& Yoon(2007)}]{vdHY2007}
van~den Heuvel, E. P.~J. \& Yoon, S.-C. 2007, Astrophysics and Space Science,
  311, 177

\bibitem[{van~der Klis(2006)}]{vdKlis2006}
van~der Klis, M. 2006, In: Compact stellar X-ray sources. Edited by Walter
  Lewin {\&} Michiel van der Klis. Cambridge Astrophysics Series, 39

\bibitem[{van Paradijs(1996)}]{vanParadijs1996}
van Paradijs, J. 1996, Astrophysical Journal Letters v.464, 464, L139

\bibitem[{Verbunt(1993)}]{Verbunt1993}
Verbunt, F. 1993, In: Annual review of astronomy and astrophysics. Vol. 31
  (A94-12726 02-90), 31, 93

\bibitem[{Webster \& Murdin(1972)}]{WM1972}
Webster, B.~L. \& Murdin, P. 1972, Nature, 235, 37

\bibitem[{Willems {et~al.}(2005)Willems, Henninger, Levin, Ivanova, Kalogera,
  McGhee, Timmes, \& Fryer}]{Willems2005}
Willems, B., Henninger, M., Levin, T., Ivanova, N., Kalogera, V., McGhee, K.,
  Timmes, F.~X., \& Fryer, C.~L. 2005, The Astrophysical Journal, 625, 324

\bibitem[{Wong {et~al.}(2014)Wong, Valsecchi, Ansari, Fragos, Glebbeek,
  Kalogera, \& McClintock}]{Wong2014}
Wong, T.-W., Valsecchi, F., Ansari, A., Fragos, T., Glebbeek, E., Kalogera, V.,
  \& McClintock, J. 2014, The Astrophysical Journal, 790, 119

\bibitem[{Wong {et~al.}(2012)Wong, Valsecchi, Fragos, \& Kalogera}]{Wong2012}
Wong, T.-W., Valsecchi, F., Fragos, T., \& Kalogera, V. 2012, The Astrophysical
  Journal, 747, 111

\bibitem[{Woosley \& Bloom(2006)}]{WB2006}
Woosley, S.~E. \& Bloom, J.~S. 2006, \araa, 44, 507

\bibitem[{Yoon {et~al.}(2006)Yoon, Langer, \& Norman}]{YLN2006}
Yoon, S.-C., Langer, N., \& Norman, C. 2006, \aap, 460, 199

\bibitem[{Zhang {et~al.}(1997)Zhang, Cui, \& Chen}]{Zhang1997}
Zhang, S.~N., Cui, W., \& Chen, W. 1997, The Astrophysical Journal, 482, L155

\bibitem[{Zhu {et~al.}(2012)Zhu, Davis, Narayan, Kulkarni, Penna, \&
  McClintock}]{Zhu2012}
Zhu, Y., Davis, S.~W., Narayan, R., Kulkarni, A.~K., Penna, R.~F., \&
  McClintock, J.~E. 2012, \mnras, 424, 2504

\bibitem[{Zurita {et~al.}(2002)Zurita, S{\'a}nchez-Fern{\'a}ndez, Casares,
  Charles, Abbott, Hakala, Rodr{\'\i}guez-Gil, Bernabei, Piccioni, Guarnieri,
  Bartolini, Masetti, Shahbaz, Castro-Tirado, \& Henden}]{Zurita2002}
Zurita, C., S{\'a}nchez-Fern{\'a}ndez, C., Casares, J., Charles, P.~A., Abbott,
  T.~M., Hakala, P., Rodr{\'\i}guez-Gil, P., Bernabei, S., Piccioni, A.,
  Guarnieri, A., Bartolini, C., Masetti, N., Shahbaz, T., Castro-Tirado, A., \&
  Henden, A. 2002, \mnras, 334, 999

\end{thebibliography}

\end{document}